\documentclass[aps,onecolumn,nofootinbib,groupedaddress,superscriptaddress,showpacs]{revtex4-1}

\usepackage{graphicx}
\usepackage{color}
\usepackage{amsmath}
\usepackage{amsfonts}
\usepackage{amssymb}
\usepackage{dcolumn}
\usepackage{hyperref}
\usepackage{enumerate}
\usepackage{bbold}
\usepackage{cancel}
\usepackage{changes}
\usepackage{mathtools}
\usepackage{booktabs}
\usepackage{cleveref}
\definecolor{green1}{RGB}{0,160,0}
\definecolor{red1}{RGB}{219,71,57}
\definecolor{orange1}{RGB}{195,106,0}
\definecolor{grey1}{RGB}{70,70,70}

\definechangesauthor[name={Emma Schepers}, color=red]{ES}
\begin{document}

\title{Incentivizing Narrow-Spectrum Antibiotic Development with Refunding}
\author{Lucas B\"ottcher}
\affiliation{Computational Medicine, UCLA, 90095-1766, Los Angeles, United States}
\email{lucasb@ucla.edu}
\affiliation{Institute for Theoretical Physics, ETH Zurich, 8093, Zurich, Switzerland}
\affiliation{Center of Economic Research, ETH Zurich, 8092, Zurich}
\email{hgersbach@ethz.ch}
\author{Hans Gersbach}
\affiliation{Center of Economic Research, ETH Zurich, 8092, Zurich}
\date{\today}
\begin{abstract}
The rapid rise of antibiotic resistance is a serious threat to global public health. Without further incentives, pharmaceutical companies have little interest in developing antibiotics, since the success probability is low and development costs are huge. The situation is exacerbated by the ``antibiotics dilemma'': Developing narrow-spectrum antibiotics against resistant bacteria is most beneficial for society, but least attractive for companies since their usage is more limited than for broad-spectrum drugs and thus sales are low.   
Starting from a general mathematical framework for the study of antibiotic-resistance dynamics with an arbitrary number of antibiotics, we identify efficient treatment protocols and introduce a market-based refunding scheme that incentivizes pharmaceutical companies to develop narrow-spectrum antibiotics: Successful companies can claim a refund from a newly established antibiotics fund that partially covers their development costs. The proposed refund involves a fixed and variable part. The latter (i) increases with the use of the new antibiotic for currently resistant strains in comparison with other newly developed antibiotics for this purpose---the resistance premium---and (ii) decreases with the use of this antibiotic for non-resistant bacteria. We outline how such a refunding scheme can solve the antibiotics dilemma and cope with various sources of uncertainty inherent in antibiotic R\&D. Finally, connecting our refunding approach to the recently established antimicrobial resistance (AMR) action fund, we discuss how the antibiotics fund can be financed. 
\end{abstract}
\maketitle
\section{Introduction}
According to the World Health Organization (WHO), antibiotic resistances are a serious threat to global public health~\cite{antibiotic_resistance_who}. Some studies see in the emergence of antimicrobial resistances (AMR) the beginning of a postantibiotic era and a threat similar to the one posed by climate change~\cite{laxminarayan2013antibiotic}. In the European Union, more than 33,000 people die every year due to infections caused by drug-resistant microbes and the corresponding yearly AMR-related healthcare costs and productivity losses are estimated to be more than 1.5 billion Euros~\cite{anderson2019averting}. 

Antibiotic resistances result from mutations in microbes and from evolutionary pressure, which selects those mutations that are resistant against certain antibiotics. The large-scale use of antibiotics in medical and agricultural settings in high-income countries led to the emergence of various multi-resistant strains. Recent findings indicate that certain strains of Enterobacteriaceae even developed resistances against the usually highly-effective class of carbapenems~\cite{jacob2013vital}. Carbapenems are so-called \emph{drugs of last resort} and only used if other antibiotic agents fail to stop the propagation of microbes.

The reasons for the development of bacterial resistances and the decline in effective treatment possibilities are multifaceted. Historically, pharmaceutical companies focused on the development of broad-spectrum antibiotics that target various strains as is the case for certain $\beta$-lactam antibiotics, whose second and third generation compounds were intentionally developed to target a broader spectrum of microbes~\cite{page2004cephalosporins,maxson2016targeted}. The rationale behind this development is that it offers pharmaceutical companies a higher return than the development of antibiotics that only target specific strains. In addition, broad-spectrum agents can be prescribed fast and without---or only with limited---diagnosis effort. On the downside, the use of broad-spectrum antibiotics seems to be correlated with an increase in antibiotic resistances~\cite{gould2011antibiotic,may2006influence,de2000antibiotic,dortch2011infection,maxson2016targeted}. Another disadvantage of broad-spectrum agents is that they are associated with outbreaks of \emph{C.~difficile} infections that result from antibiotic-induced disturbances of the gut microbiome~\cite{kelly2008clostridium,bartlett2010clostridium,maxson2016targeted}.

These  examples illustrate that the use of narrow-spectrum antibiotics may lead to a slower development of antibiotic resistance and to lower risks of \emph{C.~difficile} infections.
However, the treatment of microbial infections with narrow-spectrum agents requires efficient diagnostic techniques to quickly and precisely determine the type of bacterial strain that causes a certain infection~\cite{maxson2016targeted}; still, the clinical diagnosis of bacterial strains is often based on traditional and slow microbial-culture methods~\cite{casadevall2009case}. Only more recently, progress in the development of advanced diagnostic techniques made it possible to reduce the diagnosis time from a few days to only a few hours only~\cite{kothari2014emerging}.\footnote{
For further information on tailored antibiotic treatment approaches (i.e., personalized medicine), see also Refs.~\cite{pacheu2012mitochondrial,moser2019antibiotic}.} 

The notions of ``narrow'' and ``broad'' spectrum antibiotics are only loosely defined. Some studies distinguish between antibiotics that are applicable to Gram-positive and Gram-negative strains of microbes. The distinction between these two categories is based on the so-called Gram strain test, which categorizes bacteria into these categories according to physiological properties of their cell walls.
As in Ref.~\cite{maxson2016targeted}, we will reserve the term ``narrow-spectrum'' for antibiotics that only affect one strain or a small number of strains when given to a patient.

Treatment strategies involving narrow-spectrum antibiotics have been implemented by some northern-European countries such as Norway and Sweden~\cite{torfoss2012norwegian,molstad2008sustained}. The Norwegian strategy is based on penicillin G and aminoglycoside as initial treatment substances~\cite{torfoss2012norwegian} and it avoids broad-spectrum $\beta$-lactam antibiotics. However, further studies are necessary to better understand the influence of such narrow-spectrum treatment approaches on the population level over time.

The objective of policy is to devise strategies that help incentivizing pharmaceutical companies to focus on the development of narrow-spectrum antibiotics. According to a recent report of the European Court of Auditors~\cite{eu_auditors}, \emph{``the antimicrobials market lacks commercial incentives to develop new treatments''}. It suggests to reallocate some of the EU AMR research budget to create economic incentives for pharmaceutical companies~\cite{watson2019eu}. In the US, the Generating Antibiotic Incentives Now (GAIN) Act from 2012 pursues similar goals by \emph{``stimulating the development and approval of new antibacterial and antifungal drugs''}~\cite{fda}. 

In this paper, we develop a complementary approach by constructing a refunding scheme for successful developments of antibiotics. The development costs of new antibiotics can amount to several billion USD and the probability of a successful development might be only a few percent~\cite{dimasi2016innovation,aardal2019antibiotic}. Our refunding scheme aims at making the development of new narrow-spectrum antibiotics nevertheless  commercially viable. This can be achieved by creating an antibiotics fund and implementing a refunding scheme for it. The refunding scheme rewards companies that have successfully developed a new antibiotic (narrow or broad) as follows. A successful company can claim a refund from the antibiotics fund to partially cover its development costs. The proposed refund involves a fixed and variable part. The variable part increases with the use of the new antibiotic for \emph{currently resistant} strains in comparison with other newly developed antibiotics for the same purpose---the resistance premium---and (ii) decreases with the use of this antibiotic for non-resistant bacteria. With an appropriate choice of refunding parameters, it becomes commercially attractive to develop a narrow-spectrum antibiotic, or to switch to such an antibiotic if it becomes feasible in the R\&D process, while developing broad-spectrum antibiotics becomes less attractive. The antibiotics fund, in turn, is continuously financed by fees levied on the use of \emph{existing} antibiotics, and should be started by initial contributions from the industry and public institutions like the recent AMR Action fund.\footnote{https://amractionfund.com/, retrieved on July 13, 2020.} 

In Sec.~\ref{sec:perspective}, we provide further details on the rationale of developing incentives for the development of narrow-spectrum antibiotics. To formulate our refunding approach, we first introduce a mathematical framework of antibiotic-resistance dynamics in Sec.~\ref{sec:modeling}. Our framework is able to account for an arbitrary number of different antibiotics, whereas previous models~\cite{uecker2018antibiotic,bonhoeffer1997evaluating,levin2004cycling} only considered two to three distinct antibiotics and compared different treatment protocols such as temporal variation and combination therapy. Similar ``low-dimensional'' descriptions of antibiotic resistance have been used to study the economic problem of optimal antibiotic use~\cite{laxminarayan2001economics}. 

In Sec.~\ref{sec:narrow_broad}, we use our general framework to derive a model variant that allows us to study the ``antibiotics dilemma'': Developing narrow-spectrum antibiotics, which are only effective against specific bacterial strains, is most beneficial for society, but least attractive for pharmaceutical companies due to their limited usage and sales volumes. We couple this variant of the general antibiotics model to our refunding scheme in Sec.~\ref{sec:refunding} and illustrate how refunding can lead to better treatment protocols and a lower share of resistant strains. The refund for the development of a certain antibiotic covers part of the development costs and satisfies the following properties: (i) The refund is strongly increasing with the use of the new antibiotic for currently resistant bacteria in comparison with other newly developed antibiotics for this purpose. (ii) The refund decreases with the use of this antibiotic for non-resistant bacteria in comparison with other antibiotics used for this purpose. In Sec.~\ref{sec:refunding_general}, we outline possibilities to design refunding schemes in terms of the general antibiotic-resistance model of Sec.~\ref{sec:modeling}. We conclude our study in Sec.~\ref{sec:conclusion}
\section{Broader perspective}
\label{sec:perspective}
\begin{figure}
    \centering
    \includegraphics{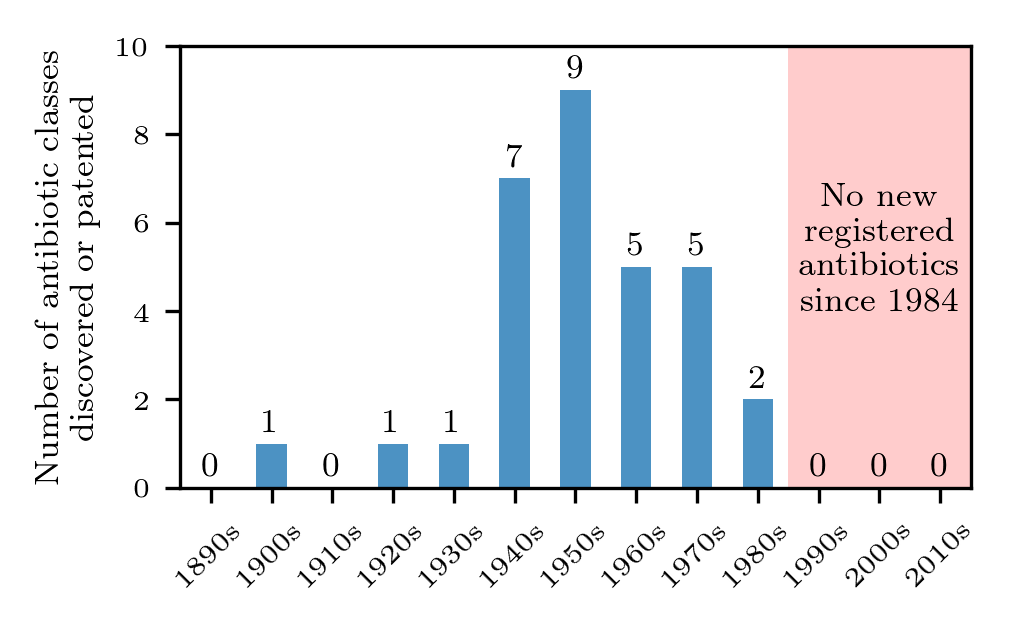}
    \caption{\textbf{Number of registered antibiotics.} We show the number of registered or patented antibiotics from 1890-2020~\cite{talkington2016scientific}.}
    \label{fig:antibiotic}
\end{figure}{}
It is useful to place our proposal into a broader context. First, incentivizing the development of narrow-spectrum antibiotics has to be matched by the development and use of efficient diagnostic techniques to quickly and precisely determine the type of bacterial strain that causes a health problem and by collecting information regarding the type of bacterial strain, the treatment and its outcome. Many OECD countries have already implemented extended reporting systems (see e.g.~the Swiss antibiotic strategy 2015~\cite{StAR}).

In addition to the development of narrow-spectrum agents, it is important to also consider alternative approaches such as medication that sustains and boosts the human immune system during infections, or improved sterilization and sanitation in hospitals~\cite{maxson2016targeted}.
Other strategies for fighting bacterial infections, such as targeting virulence or treatment with antibodies or phage~\cite{kutateladze2008phage,altamirano2019phage,kortright2019phage}, are also alternatives to antibiotics.

In practice, it will not be easy to encourage pharmaceutical companies to refocus their R\&D activities. The disappointing finding that genomics did not lead to many new classes of antibiotics caused the close-down of many antibiotic research laboratories~\cite{coates2011novel,laxminarayan2013antibiotic}. Currently, the pipeline of new ideas seems to be rather small (see Fig.~\ref{fig:antibiotic}), while the costs of clinical trials are very high. The authors of Ref.~\cite{martin2017much} analyzed the clinical trial costs of 726 studies that were conducted between 2010-2015. In the initial clinical trial phase, the median cost was found to be 3.4 million USD and the median cost of phase III\footnote{Phase III clinical trials are the last phase of clinical research that has to be satisfactorily completed before regulatory agencies will approve a new drug. Such trials usually involve large patient groups (ca.~300--3000 volunteers who have the disease or condition) and require comparatively long observation periods ranging between 1--4 years~\cite{fda_phaseIII}.} in the development process was reported to be more than 20 million USD. High development costs of antibiotic drugs limit the number of players in this area and require major companies to be involved in the development process. A good research ecosystem for antibiotic development necessarily involves large companies, entailing  significant in-house efforts, but also collaborations with academia, buying or investing in SMEs, and joint ventures with other large pharmaceutical companies. An appropriately designed refunding scheme can help to foster such a research ecosystem.

There are also arguments that the development of new antibiotics is not so critical. The emergence of antibiotic resistance, even for new classes of antibiotics, is inevitable and one may conclude that research and development efforts should mainly focus on antibiotic substances that are effective against highly resistant strains (see e.g.~Ref.~\cite{laxminarayan2013antibiotic}).

A recent report published by the European Observatory on Health Systems and Policies~\cite{anderson2019averting} suggests a multifold R\&D approach to combat AMR that includes: (i) push incentives (e.g., direct funding and tax incentives) and pull incentives (e.g., milestone prize and patent buyout) for the development of new antibiotics, (ii) research in diagnostics (e.g., rapid tests to distinguish between bacterial and viral infections), and (iii) vaccine research. 

Our proposed refunding scheme is a complementary public-private initiative to foster the development of new antibiotics.
\section{Modeling antibiotic treatment}
\label{sec:modeling}
\begin{figure}
\centering
\includegraphics[width=0.65\textwidth]{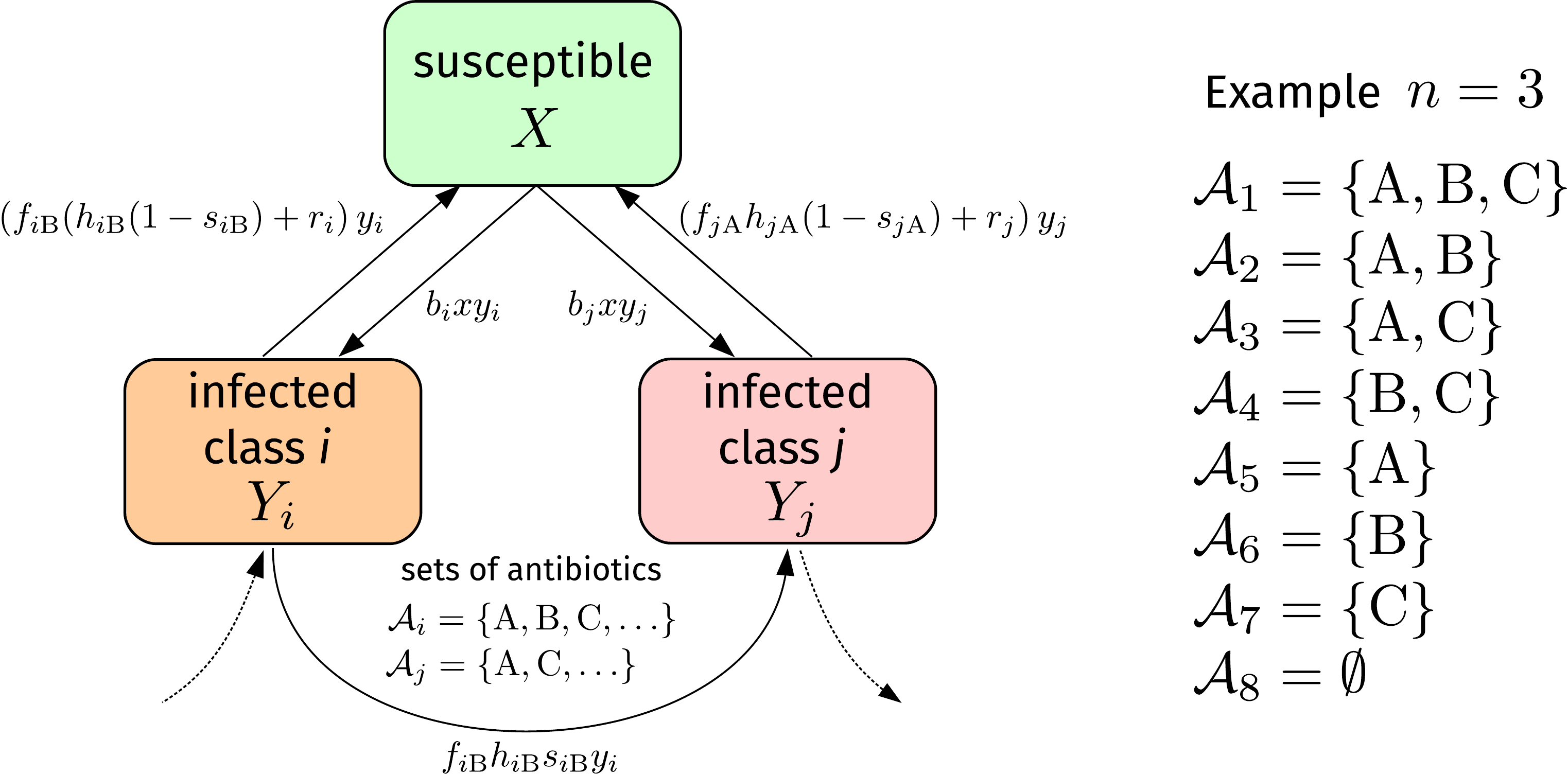}
\caption{\textbf{Model schematic.} Susceptible individuals ($X$) can be infected with bacterial strain $i$ at rate $b_i$. Infected individuals in state $Y_i$ recover spontaneously at rate $r_i$. The corresponding antibiotic-induced recovery rate of antibiotic $\rm{A}$ is $f_{i\rm{B}} h_{i\rm{B}}$. However, only a fraction $1-s_{i\rm{B}}$ of individuals in state $Y_i$ recovers after a treatment with antibiotic $\rm{B}$. The remaining fraction $s_{i\rm{B}}$ becomes resistant against antibiotic $\rm{B}$ and ends up in a compartment $Y_j$ of bacterial strains exhibiting more resistances. The sets of effective antibiotics in compartments $Y_i$ and $Y_j$ are $\mathcal{A}_i$ and $\mathcal{A}_j$, respectively. Infection and recovery processes with the respective rates are also present in compartment $Y_j$. For the case of $n=3$ antibiotics, we show the possible antibiotic-treatment classes $\mathcal{A}_1,\mathcal{A}_2,\dots,\mathcal{A}_8$.}
\label{fig:model}
\end{figure}
\subsection{Treatment of infections with $n$ antibiotics}
In order to motivate the refunding scheme and model population-level dynamics of antibiotic resistance, we first introduce a corresponding mathematical framework that is able to account for an arbitrary number of antibiotics. We describe the interaction between infectious and susceptible individuals in terms of a susceptible-infected-susceptible (SIS) model~\cite{keeling2011modeling}
whose infected compartment is sub-divided into compartments that can each be treated with certain antibiotics. We indicate a susceptible state by $X$ and use $Y_i$ to denote infected states that are sensitive to antibiotics in the set $\mathcal{A}_i$. If the set $\mathcal{A}_2$ contains two antibiotics $\rm{A}$ and $\rm{B}$ (i.e., $\mathcal{A}_2=\{\mathrm{A},\mathrm{B}\}$), individuals in state $Y_2$ can be treated with these two antibiotics but not with a potentially available third antibiotic $\rm{C}$ that can be used to treat individuals in state $Y_1$, where $\mathcal{A}_1=\{\mathrm{A},\mathrm{B},\mathrm{C}\}$ (see Fig.~\ref{fig:model}). 

We describe antibiotic-resistance dynamics in terms of a mass-action model of multiple antibiotic therapy (see Fig.~\ref{fig:model}):
\begin{align}
\begin{split}
\frac{\mathrm{d} x}{\mathrm{d} t}&=-x\sum_{i=1}^N b_i y_i+\sum_{i=1}^{N}r_i y_i + \left[\sum_{i=1}^N \sum_{j\in \mathcal{A}_i} f_{i j} h_{ij} \left(1-s_{ij}\right) y_i\right]+\lambda - d x\,, \\
\frac{\mathrm{d} y_i}{\mathrm{d} t}&=b_i x y_i - r_i y_i-c_i y_i - \sum_{j\in \mathcal{A}_i} f_{i j} h_{ij} y_i+\sum_{k < i} \sum_{j\in \mathcal{S}(\mathcal{A}_i)} f_{kj} h_{kj} s_{kj} y_k\,,
\end{split}
\label{eq:rate1}
\end{align}
where $x$ and $y_i$ are the fractions of individuals in states $X$ and $Y_i$, respectively. An infection with bacterial strain $i$ occurs at rate $b_i$ and the corresponding (spontaneous) recovery rate is $r_i$. Throughout the manuscript, we use the convention that as $i$ increases, the corresponding bacteria become more resistant. We also account for the fitness cost associated with antibiotic resistance (i.e., $r_{i}-r_{1} >0$ for all $i \in \{2,\dots,N\}$)~\cite{andersson2006biological}. Additional antibiotic-induced recovery from compartment $i$ with antibiotic $j\in \mathcal{A}_i$ occurs with rate $f_{i j} h_{ij}$, where $f_{i j}$ is the proportion of antibiotic $j\in \mathcal{A}_i$, relative to other antibiotics, that is used to treat $Y_i$. However, only a fraction $1-s_{ij}$ actually recovers, whereas the remaining fraction $s_{ij}$ develops a resistance against antibiotic $j\in \mathcal{A}_i$. The birth rate of new susceptible individuals is $\lambda$ and the corresponding death rate is $d$. For infected individuals in state $Y_i$, the death rate is $c_i$. The set $\mathcal{S}(\mathcal{A}_i)$ contains all antibiotics that were used to arrive at the partially or completely resistant compartment $Y_i$ from other states $Y_k$ ($k<i$) with less resistances.
\begin{figure}
\includegraphics[width=0.6\textwidth]{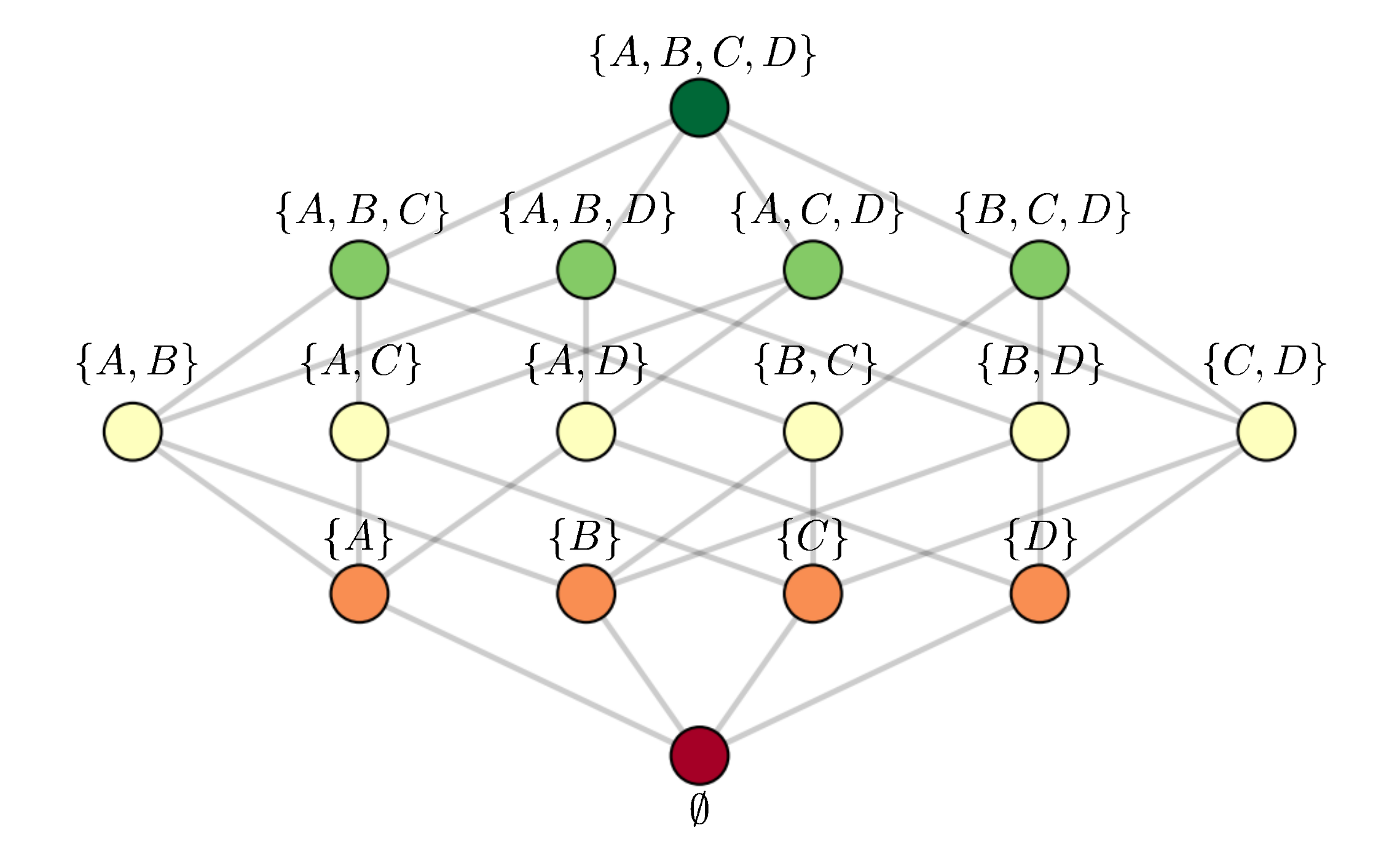}
\caption{\textbf{Antibiotic-resistance network.} For the case of $n=4$ antibiotics, we show the antibiotic-resistance network. Nodes correspond to states in which the indicated antibiotics are effective and edges between nodes represent the development of resistant strains due to the usage of certain antibiotics. In the shown example, single-antibiotic therapy is being used. That is, no combinations of antibiotics are being administered to patients.}
\label{fig:mutation_tree}
\end{figure}
For example, the use of single antibiotics (i.e., one per patient) is described by
\begin{equation}
\mathcal{S}(\mathcal{A}_i)=\left\{\bigcup_{k} \mathcal{A}_k\setminus \mathcal{A}_i \biggr\vert \mathcal{A}_k\in\{\mathcal{A}_1,\mathcal{A}_2,\dots,\mathcal{A}_N\},|\mathcal{A}_k|=|\mathcal{A}_i|+1,\mathcal{A}_k \cap \mathcal{A}_i\neq \emptyset \right\}
\label{eq:set_Si}
\end{equation}
and, in this case, each antibiotic is used with proportions $f_{i j} = 1/|\mathcal{A}_i|$, where $|\mathcal{A}|$ is the cardinality of set $\mathcal{A}$. We illustrate an example of a corresponding antibiotic-resistance network for $n=4$ antibiotics in Fig.~\ref{fig:mutation_tree}. Nodes in such a resistance network represent states $Y_i$ and edges describe treatment strategies. In the example we show in Fig.~\ref{fig:mutation_tree}, only single antibiotics (no combinations) are being used for treatment.

The general antibiotic-resistance model (see Eq.~\eqref{eq:rate1}) has $N$ different compartments, which correspond to $N$ resistance states, each accounting for a certain set of effective antibiotics. We denote the total number of antibiotics by $n$. What is the number of resistance states $N$ that belongs to a certain number of antibiotics $n$? Considering the antibiotic-resistance network of Fig.~\ref{fig:mutation_tree}, we observe that the total number of resistance states $N$ is the sum over all possible combinations of single antibiotics plus one (representing the completely resistant state). For $n$ different antibiotics, we thus have to consider $N=1+\sum_{k=1}^n \binom{n}{k}=2^n$ different states $Y_i$ ($i\in\{1,2,\dots,N\}$). We order them in the following way. We denote by $Y_1$ the infected state that can be successfully treated with all antibiotics, whereas $Y_N$ represents an infection with a completely resistant strain. Let $k\leq n$ be the number of effective antibiotics. For a wild-type strain, the number of effective antibiotics is $k=n$. In each layer of the antibiotic-resistance network, there are $\binom{n}{k}$ different strains. For a treatment with single antibiotics (see Fig.~\ref{fig:mutation_tree}), there are always $\binom{n}{k}$ nodes with $k$ edges in a certain layer that need to be connected to $\binom{n}{k-1}$ nodes in the following layer. The relative difference is
\begin{equation}
\frac{k \binom{n}{k}}{\binom{n}{k-1}}=\frac{(n-k+1)\,!}{(n-k)\,!}=n-k+1\,.
\end{equation}
This equation implies that $n-k+1$ elements from the current layer are mapped to one element in the next layer. In the first layer ($k=4$) of the network that we show in Fig.~\ref{fig:mutation_tree}, one element from the current layer is mapped to one element in the next layer. In the second layer ($k=3$), two elements are mapped to one element in the third layer. Similar considerations apply to other treatment protocols (e.g., combination treatment with multiple antibiotics) and help to formulate the corresponding set of rate equations.

Previous models only considered the treatment with two and three antibiotics~\cite{bonhoeffer1997evaluating,levin2004cycling,uecker2018antibiotic}. Our generalization to $N$ compartments enables us to provide insights into the higher-dimensional nature of the dynamical development of antibiotic resistances. In Appendices \ref{app:comparison} and \ref{app:general} we compare the outlined single-antibiotic therapy approach with combination treatment for different numbers of antibiotics. We also demonstrate in Appendix \ref{app:general} that the mathematical form of the stationary solution of Eq.~\eqref{eq:rate1} is unaffected by the number of antibiotics. Still, more antibiotics can be useful to slow down the development of completely resistant strains, suggesting that rolling out more antibiotics is useful. However, as we will discuss below, fostering the development of particular types of narrow-spectrum antibiotics is much more powerful to slow down the occurrence of completely resistant strains and reduce the number of deaths than developing broad-spectrum antibiotics. 
\subsection{Performance measures}
We can compare different treatment protocols in terms of different metrics including the total stationary population
\begin{equation}
P^\ast \coloneqq x^\ast+\sum_{i=1}^N y_i^\ast\,,
\label{eq:stat_pop}
\end{equation}
where the asterisk denotes the stationary densities of $x$ and $y_i$. Another possible metric is the gain of healthy individuals
\begin{equation}
G(T)\coloneqq\int_0^{T} x(t)\, \mathrm{d}t-\int_0^{T}x(t;h_{ij}=0) \mathrm{d}t\,,
\label{eq:gain}
\end{equation}
through antibiotic treatment during some time, denoted by $T$, where $x(t;h_{ij}=0)$ denotes the proportion of susceptible individuals in the absence of treatment (i.e., $h_{ij}=0$ for all $i,j$). 

Finally, we can calculate the time until half of the infected individuals are infected by bacterial strains that are resistant against any antibiotic. This ``half-life'' of non-resistance is given by:  
\begin{equation}
T_{1/2}\coloneqq \left\{t\bigg|\frac{y_N(t)}{\sum_{i=1}^{N} y_i(t)}=\frac{1}{2}\right\}\,.
\label{eq:T12}
\end{equation}
As mentioned above and proven in the Appendix, the long-term stationary population $P^\ast$ is no suitable performance measure, since $P^\ast$ is identical for all treatment protocols that we will consider in the following sections. However, both $G(T)$ and $T_{1/2}$ are suitable measures to compare different development strategies for antibiotics. In addition, we will also use $G_{1/2}\coloneqq G(T_{1/2})$ as performance metric.
\section{Narrow versus broad spectrum antibiotics}
\label{sec:narrow_broad}
\subsection {Research and development opportunities}
\label{sec:development_oppur}
To provide a formal representation of both the antibiotics dilemma and the refunding scheme, we consider the simplest case with $n=2$ antibiotics. In Sec.~\ref{sec:refunding_general}, we discuss a more general refunding approach that can be used in conjunction with the general antibiotic-resistance model of Sec.~\ref{sec:modeling}.

For the model variant with $n=2$ antibiotics, the corresponding sets of antibiotics for the $N=2^2=4$ infected compartments are $\mathcal{A}_1=\{\mathrm{A},\mathrm{B}\}$, $\mathcal{A}_2=\{\mathrm{A}\}$, $\mathcal{A}_3=\{\mathrm{B}\}$, and $\mathcal{A}_4=\emptyset$.

We assume that antibiotic $A$ is already on the market. For the development of a second antibiotic, there are two possibilities for pharmaceutical companies.

\begin{itemize}
\item Antibiotic $\mathrm{B}_1$: This is a broad-spectrum antibiotic that is as effective as antibiotic $\rm{A}$ against wild-type strains. It is also effective against strains that are resistant against $\rm{A}$.
\item Antibiotic $\mathrm{B}_2$: This is a narrow-spectrum antibiotic that is, by a factor $1-\alpha_{\mathrm{B}}$ ($0\leq \alpha_{\mathrm{B}} \leq 1$), less effective against wild-type strains. However, antibiotic $B_2$ is, by a factor $1+\gamma_\mathrm{B}$ ($ \gamma_{\mathrm{B}} >0$), more effective against strains that are resistant against $\rm{A}$ than $\rm{A}$ is effective against strains that are resistant to $\mathrm{B}_2$.
\end{itemize} 

Note that we use the terms ``narrow'' and ``broad'' to classify antibiotics according to their effectiveness against certain bacterial strains \footnote{An alternative mathematical definition of ``narrow'' and ``broad'' would be to classify combination treatment as a broad-spectrum approach and single-antibiotic therapy as narrow.}.

We will later turn to costs and chances to develop such antibiotics, but at the moment, we consider what happens if either $\mathrm{B}_1$ or $\mathrm{B}_2$ is being developed and used for treating patients. 
\subsection{The model}
\begin{figure}
\centering
\includegraphics[width=0.49\textwidth]{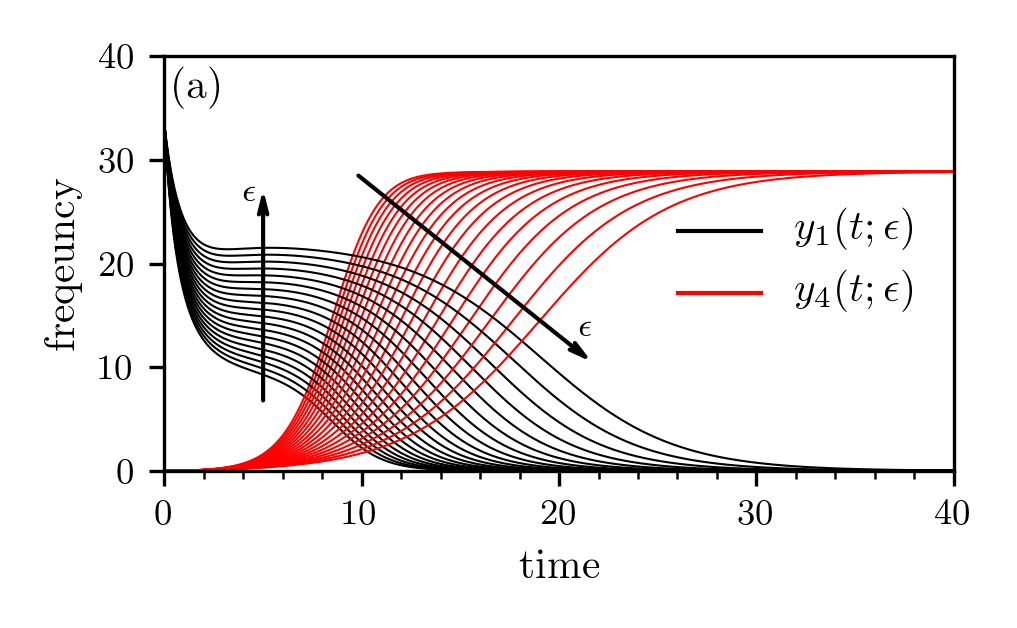}
\includegraphics[width=0.49\textwidth]{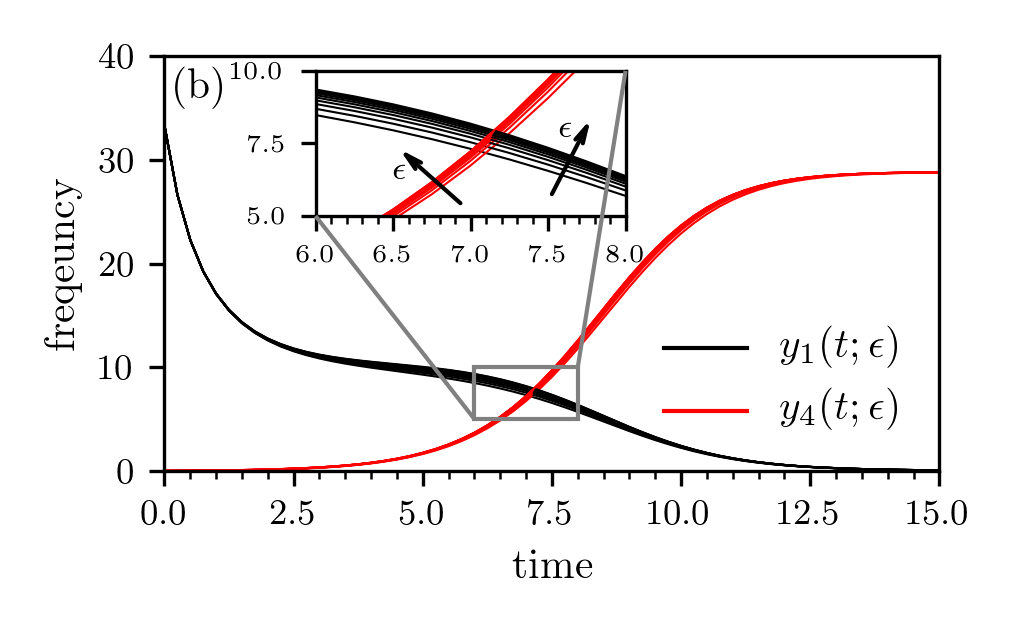}
\caption{\textbf{Growth of multi-resistant strains under 50/50 and 100/0 treatment.} The evolution of wild-type $y_1$ (black solid line) and completely resistant infections $y_4$ (red solid line) under 50/50 treatment with $f_{1\mathrm{A}}=f_{1\mathrm{B}_1}=0.5$ (a) and 100/0 treatment with $f_{1\mathrm{A}}=1$ and $f_{1\mathrm{B}_2}=0$ (b). To obtain the solutions shown, we numerically solve Eqs.~\eqref{eq:narrow_broad_new} with a classical Runge-Kutta scheme in the time interval $[0,T]$ with $T=100$ setting $\lambda=100$, $d=1$, $c=1.5$, $b=0.03$, $r_{i}=(2-k) 0.1$ ($k$ is the number of effective antibiotics in the respective layer), $h=1$, $s=0.05$, $\alpha_\mathrm{B}=\gamma_{\mathrm{B}}=\epsilon \geq 0 $. The initial conditions are $x(0)=50$, $y_1(0)=33.33$, $y_2(0)=y_3(0)=y_4(0)=0$.}
\label{fig:y1y4}
\end{figure}
For both types of antibiotics, we derive the corresponding population-level dynamics according to Eq.~\eqref{eq:rate1}:
\begin{align}
\begin{split}
\frac{\mathrm{d} x}{\mathrm{d} t}&=-b x \left( y_1+y_2+y_3+y_4\right)+r_1 y_1+r_2 y_2 + r_3 y_3+r_4 y_4 \\
&+ h (1-s) \left\{y_1\left[f_{1\mathrm{A}}+(1-\epsilon)f_{1\mathrm{B}}\right]+y_2+y_3(1+\epsilon)\right\}+\lambda - d x\,,\\
\frac{\mathrm{d} y_1}{\mathrm{d} t}&=\left[b x  - r_1 -c-h\left(f_{1\mathrm{A}}+(1-\epsilon) f_{1\mathrm{B}}\right)\right]y_1\,, \\
\frac{\mathrm{d} y_2}{\mathrm{d} t}&=\left(b x - r_2 -c-h\right)y_2+ h s (1-\epsilon) f_{1\mathrm{B}} y_1\,, \\
\frac{\mathrm{d} y_3}{\mathrm{d} t}&=\left( b x - r_3-c-h (1+\epsilon) \right)y_3+ h s f_{1 \mathrm{A}} y_1\,, \\
\frac{\mathrm{d} y_4}{\mathrm{d} t}&=\left(b x - r_4-c\right) y_4 + h s\left[ y_2+(1+\epsilon)y_3 \right] \,,
\end{split}
\label{eq:narrow_broad_new}
\end{align}
where we have set $\gamma_{\mathrm{B}}=\alpha_{\mathrm{B}}=\epsilon \in [0,1]$ (see Sec.~\ref{sec:development_oppur}). Furthermore, we assumed that the antibiotic-induced recovery rate, the infection rate, and the fraction of individuals that develop an antibiotic resistance is constant for all antibiotics, i.e.~$h_{ij}=h$, $b_i=b$ and $s_{ij}=s$. For modeling antibiotic $\mathrm{B}_1$, we simply set $\epsilon=0$. If antibiotic $\mathrm{B}_2$ is present, we set the values of these parameters to the efficiency disadvantages and advantages of $\mathrm{B}_2$ relative to $\mathrm{A}$. Note that the assumption of an equal infection rate $b$ of different strains is justified by corresponding empirical findings~\cite{chehrazi2019dynamics}.
We now focus on four different treatment strategies:
\begin{itemize}
\item[I.] Treatment with antibiotic $\mathrm{B}_1$ and symmetric use of antibiotics (i.e., 50/50):\\ 
Hence, we have $\epsilon=0$. Moreover, since we consider a symmetric use, $f_{1 \rm{A}}=0.5$ and $f_{1 \rm{B}}=0.5$, this case describes a treatment strategy where 50\% of patients with a wild-type-strain infection receive antibiotic A and the remaining 50\% receive antibiotic $\mathrm{B}_1$. Both antibiotics $\rm{A}$ and $\mathrm{B}_1$ have the same effect on strains $1$ and $2$ and $1$ and $3$, respectively.

\item[II.] Treatment with antibiotic $\mathrm{B}_1$ and asymmetric use of antibiotics (i.e., 100/0):\\ 
Hence, we again have $\epsilon=0$. We only use the new antibiotic $\mathrm{B}_1$ against strains that are resistant against $\rm{A}$. All patients with a wild-type-strain infection receive antibiotic A, i.e. $f_{1 \rm{A}}=1$ and $f_{1 \rm{B}}=0$. 

\item[III.] Treatment with antibiotic $\mathrm{B}_2$ and symmetric use of antibiotics:\\
The symmetric use implies 
$f_{1 \rm{A}}/f_{1 \rm{B}}=50/50$ for wild-type-strain infections. The prefactor $1-\epsilon$ accounts for the corresponding recovery-rate difference in the wild-type compartment. However, antibiotic $\mathrm{B}_2$ is more effective in compartment $3$, where individuals have an antibiotic-induced recovery rate of $h (1+\epsilon)$. 

\item[IV.] Treatment with antibiotic $\mathrm{B}_2$ and asymmetric use of antibiotics:\\
In this scenario, we only use antibiotic $\rm{A}$ in compartment $Y_1$ and thus set $f_{1 \rm{A}} = 1$ and $f_{1 \rm{B}}=0$. 

\end{itemize}
\subsection {Comparisons}
\begin{figure}
\centering
\includegraphics[width=0.49\textwidth]{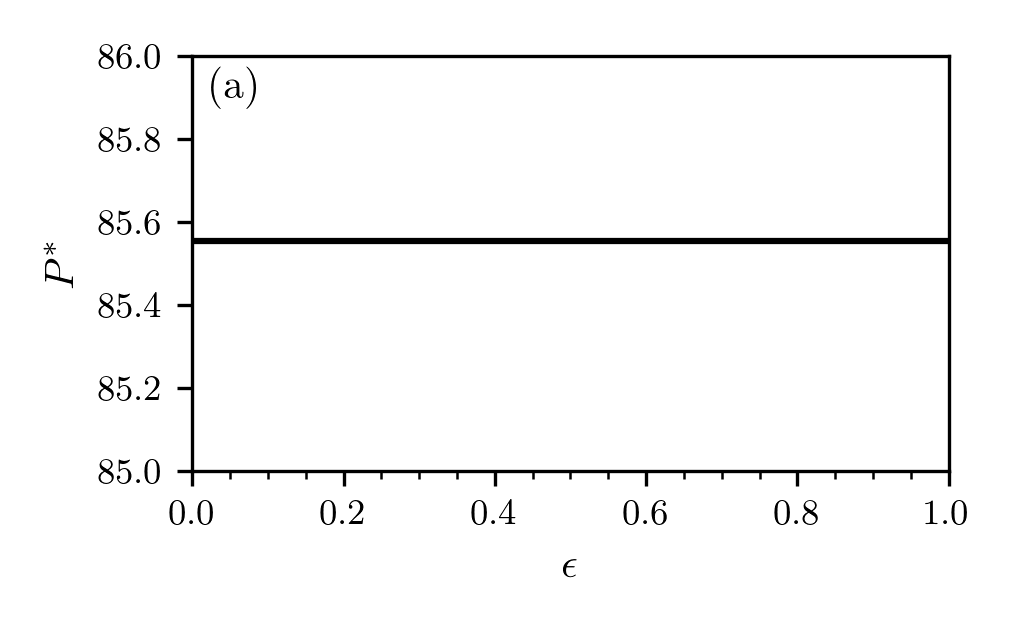}
\includegraphics[width=0.49\textwidth]{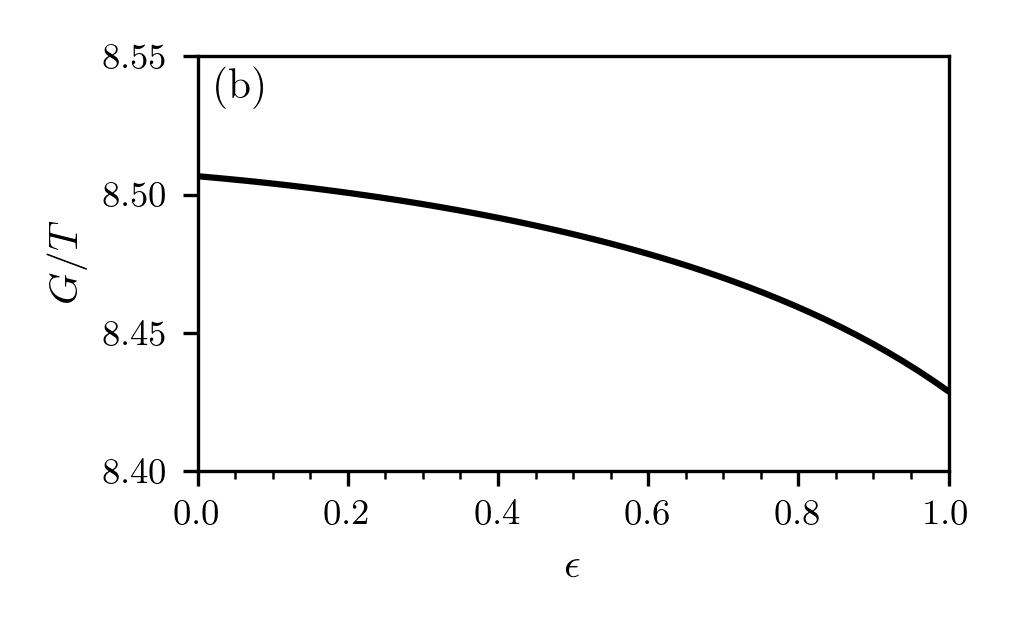}
\includegraphics[width=0.49\textwidth]{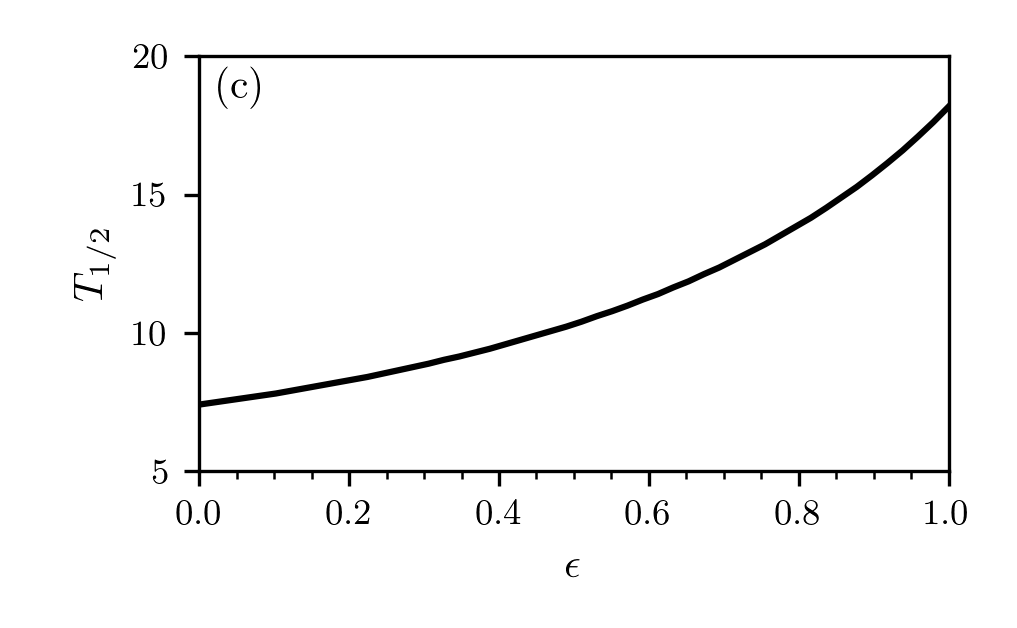}
\includegraphics[width=0.49\textwidth]{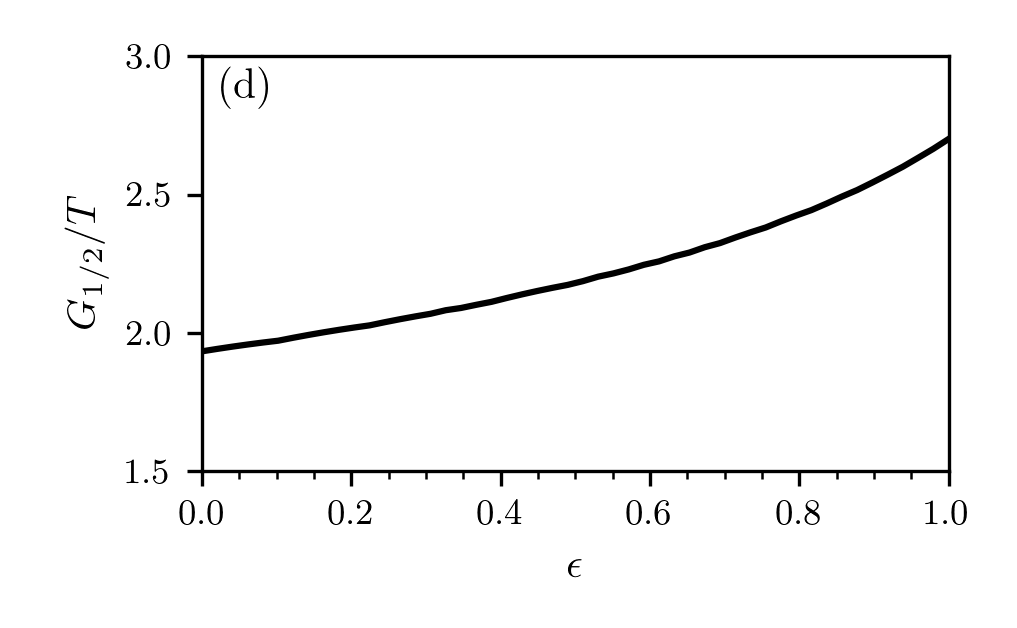}
\caption{\textbf{Treatment with two antibiotics (50/50 strategy).} We numerically solve Eqs.~\eqref{eq:narrow_broad_new} with a classical Runge-Kutta scheme in the time interval $[0,T]$ with $T=100$ setting $\lambda=100$, $d=1$, $c=1.5$, $b=0.03$, $r_{i}=(2-k) 0.1$ ($k$ is the number of effective antibiotics in the respective layer), $h=1$, $s=0.05$, and $f_{1\mathrm{A}}=f_{1\mathrm{B}_1}=0.5$, $\alpha_\mathrm{B}=\gamma_{\mathrm{B}}=\epsilon \geq 0 $. The initial conditions are $x(0)=50$, $y_1(0)=33.33$, $y_2(0)=y_3(0)=y_4(0)=0$.}
\label{fig:simulations50_50}
\end{figure}
\begin{figure}
\centering
\includegraphics[width=0.49\textwidth]{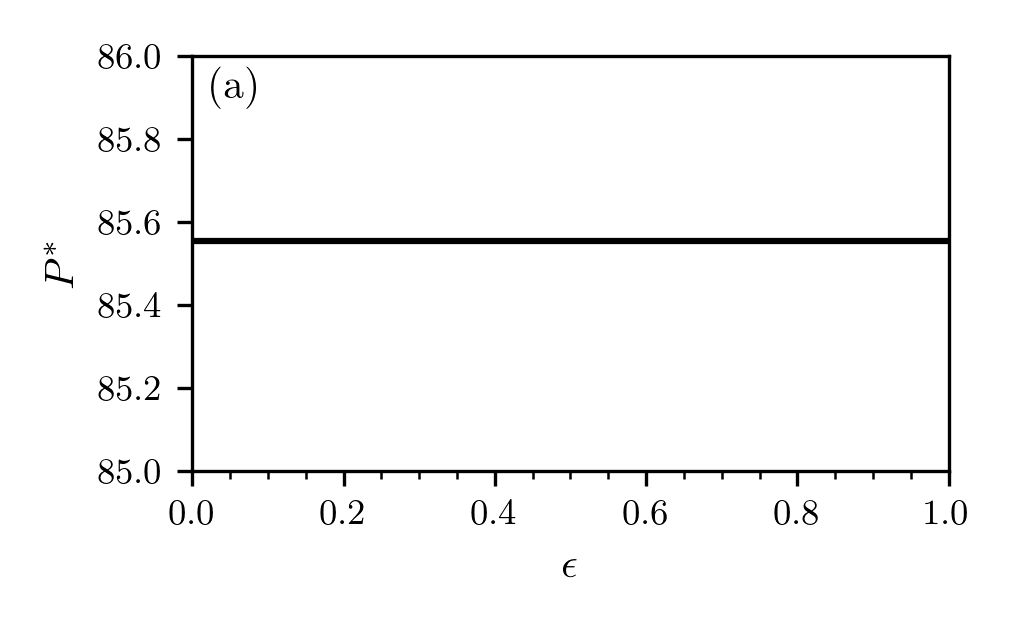}
\includegraphics[width=0.49\textwidth]{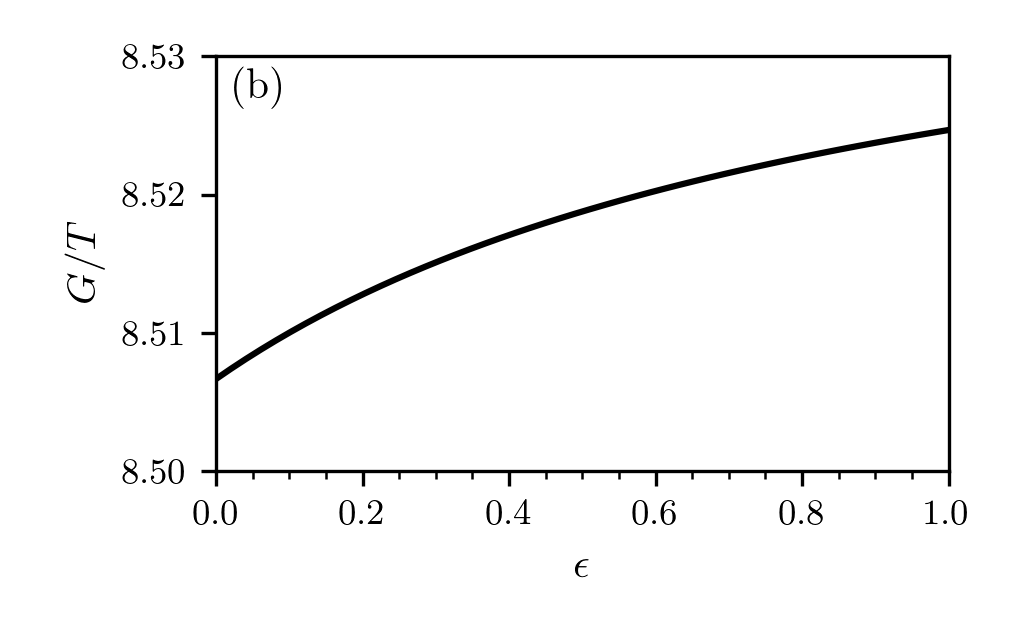}
\includegraphics[width=0.49\textwidth]{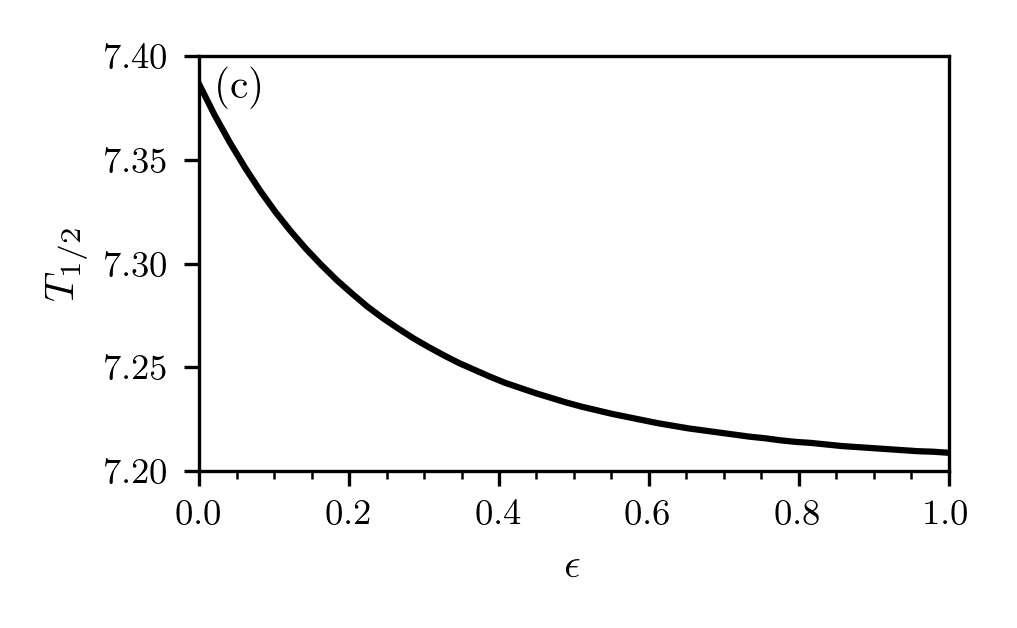}
\includegraphics[width=0.49\textwidth]{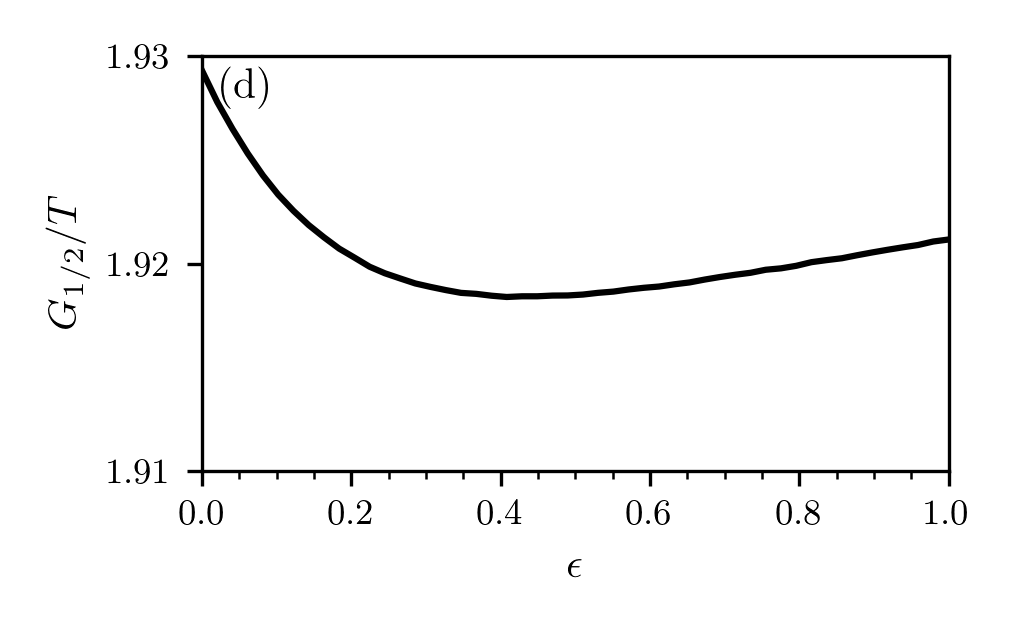}
\caption{\textbf{Treatment with two antibiotics (100/0 strategy).} We numerically solve Eqs.~\eqref{eq:narrow_broad_new} with a classical Runge-Kutta scheme in the time interval $[0,T]$ with $T=100$ setting $\lambda=100$, $d=1$, $c=1.5$, $b=0.03$, $r_{i}=(2-k) 0.1$ ($k$ is the number of effective antibiotics in the respective layer), $h=1$, $s=0.05$, and $f_{1\mathrm{A}}=1$, $f_{1\mathrm{B}_2}=0$, $\alpha_\mathrm{B}=\gamma_{\mathrm{B}}=\epsilon \geq 0$. The initial conditions are $x(0)=50$, $y_1(0)=33.33$, $y_2(0)=y_3(0)=y_4(0)=0$.}
\label{fig:simulations100_0}
\end{figure}
\begin{figure}
\centering
\includegraphics[width=0.49\textwidth]{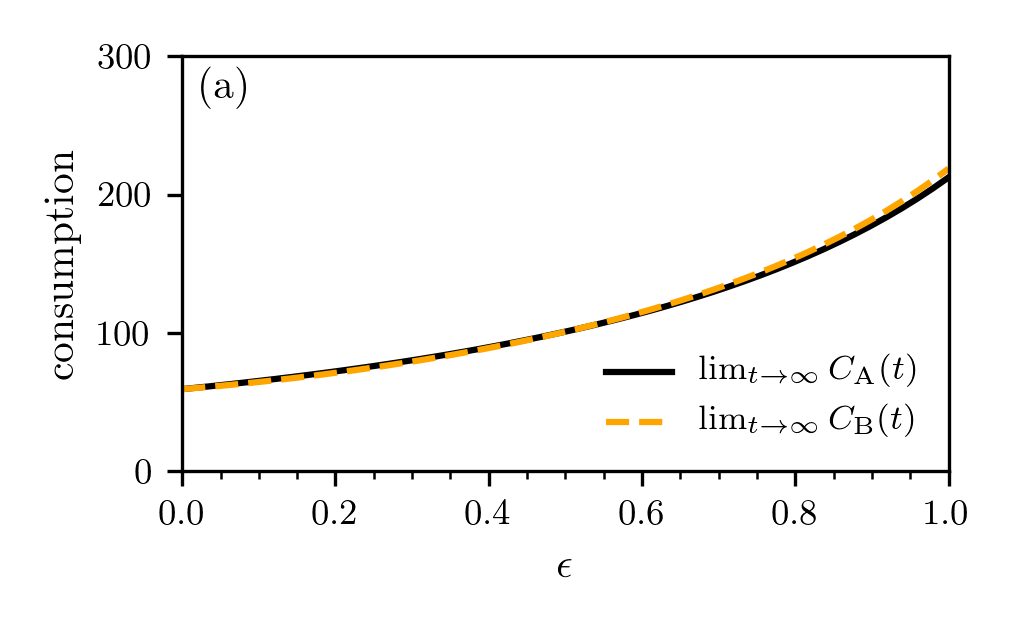}
\includegraphics[width=0.49\textwidth]{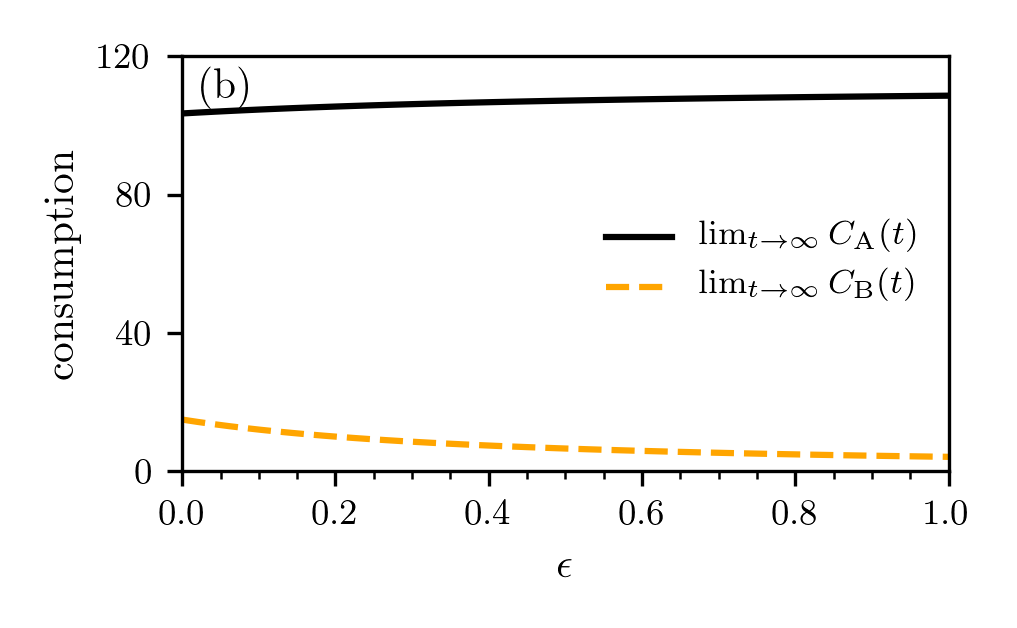}
\caption{\textbf{Antibiotic consumption.} We show the stationary antibiotic consumption $ \lim_{t \rightarrow \infty}C_{\rm A}$ and $\lim_{t \rightarrow \infty} C_{\rm B}$ for the 50/50 treatment with $f_{1\mathrm{A}}=f_{1\mathrm{B}_1}=0.5$ (a) and 100/0 treatment with $f_{1\mathrm{A}}=1$ and $f_{1\mathrm{B}_2}=0$ (b). We numerically solve Eqs.~\eqref{eq:narrow_broad_new} with a classical Runge-Kutta scheme in the time interval $[0,T]$ with $T=100$ setting $\lambda=100$, $d=1$, $c=1.5$, $b=0.03$, $r_{i}=(2-k) 0.1$ ($k$ is the number of effective antibiotics in the respective layer), $h=1$, $s=0.05$, and $\alpha_\mathrm{B}=\gamma_{\mathrm{B}}=\epsilon \geq 0$. The initial conditions are $x(0)=50$, $y_1(0)=33.33$, $y_2(0)=y_3(0)=y_4(0)=0$.}
\label{fig:consumption}
\end{figure}
We now compare treatment strategies I--IV in terms of $P^*$, $G$, $T_{1/2}$, $G_{1/2}$ (see Figs.~\ref{fig:simulations50_50} and \ref{fig:simulations100_0}), and the use of the antibiotics A and $\mathrm{B}_1$ and $\mathrm{B}_2$, respectively. We can keep track of the consumption of antibiotics $\rm A$ and $\rm B$ by integrating
\begin{align}
\frac{\mathrm{d} C_{\rm A}}{\mathrm{d} t} = f_{1 \rm A} y_1 + y_2\quad \text{and} \quad  \frac{\mathrm{d} C_{\rm B}}{\mathrm{d} t} = f_{1 \rm B} y_1 + y_3
\end{align}
over time. We now compare scenarios I--IV, by varying $\epsilon$ from 0 to 1.

We first look for differences between the evolution of the wild-type and fully resistant compartments under 50/50 and 100/0 treatment. Based on the simulation data that we show in Fig.~\ref{fig:y1y4}, we conclude that the variability in $y_1$ and $y_4$ is much larger under 50/50 treatment than under 100/0 treatment. Comparing the performance measures $P^*$, $G$, $T_{1/2}$, and $G_{1/2}$ (see Figs.~\ref{fig:simulations50_50} and \ref{fig:simulations100_0}), we find that the $50/50$ and $100/0$ are equivalent for $\epsilon=0$. For larger values of $\epsilon$, the gain $G$ of the $50/50$ strategy is smaller than the gain of the $100/0$ strategy. Differences between the two treatment strategies are also reflected in the final consumption $C_{\rm A}$ and $C_{\rm B}$ of antibiotics A and B (see Fig.~\ref{fig:consumption}). For $\epsilon \gtrsim 0.5$, the $100/0$ treatment leads to a significantly lower consumption/return of both antibiotics compared to $50/50$ protocol.

These figures highlight a fundamental dilemma. Developing a narrow-spectrum antibiotic $\mathrm{B}_2$ is highly beneficial for society, but then it should only be used very little, namely against the strains which are resistant against antibiotic A. We refer to this issue as the \emph{antibiotics dilemma}: Developing a narrow-spectrum antibiotic against resistant bacteria is most attractive for society, but least attractive for companies, since usage is should be limited, so that sales are low.
\section{Refunding schemes}
\label{sec:refunding}
\subsection{The basic principles}

The situation is further complicated by additional properties of antibiotics development and usage. First, the development costs are enormous, in the range of several billion USD, and second the chances to succeed are low. This is true in general for new drugs~\cite{dimasi2016innovation} but more pronounced for antibiotics, where the success probability may be as low as 5\%~\cite{aardal2019antibiotic}. Third, once a narrow-spectrum antibiotics is developed and it is also effective to some degree against wild-type strains, using it for wild-type strains should not be commercially attractive. 

To overcome the antibiotics dilemma and associated complications, we suggest to introduce a refunding scheme for the use of newly developed antibiotics. The refunding scheme works as follows: 

\begin{enumerate}
\item An antibiotics fund should be started by initial contributions from the industry and public institutions like the establishment of the recent AMR Action fund. In addition all antibiotic use is charged with a small fee which is channeled continuously into the antibiotics fund. 

\item Firms that develop new antibiotics obtain a refund from the fund.

\item The refund for a particular antibiotic is calculated with a formula that satisfies the following three properties:
\begin{itemize}
\item There is a fixed payment for a successful development of an antibiotic, i.e. an antibiotic that is approved by the public health agency responsible for such approvals (e.g., the U.S.~Food and Drug Administration (FDA)). This part is in the spirit of Ref.~\cite{kremer1998patent}, as it is equivalent to an advanced market commitment. Pharmaceutical companies know that once an approved patent for a new antibiotics is awarded, they will be reimbursed part of their development costs. 
\item  The refund is strongly increasing with the use of the new antibiotic for currently resistant bacteria, compared to other newly developed antibiotics for this purpose. This part is the \emph{resistance premium}.  
\item The refund is weakly or strongly declining in the use of the antibiotics for non-resistant bacteria, compared to other antibiotics used for this purpose.
\end{itemize}
\end{enumerate}

The objective of our refunding scheme is to financially incentivize pharmaceutical companies to undertake R\&D for new narrow-spectrum antibiotics, by using a minimum-size antibiotics fund. As we will demonstrate below, all three elements are necessary to achieve this purpose.

Several remarks are in order: First, refunding schemes are widely discussed in the environmental literature to provide incentives for firms to reduce pollution~\cite{gersbach2012global}. Second, simple forms of refunding schemes could also be used in other contexts where pharmaceutical companies have only little financial interest to investing in drug research, due to potentially low sales volumes. This is, for instance, the case for orphan drug development and vaccine research for viral infections including SARS and Ebola or enduring epidemic diseases in the sense of Ref.~\cite{bell2009macroeconomics}. However, for such cases, refunding schemes are much easier to construct, since they can solely rely on the usage, e.g.\ the number of vaccinated individuals. For antibiotics---because of the antibiotics dilemma---one has to construct new types of refunding schemes with ``sticks and carrots'': The carrot for using the antibiotic against bacterial strains resistant against other antibiotics and sticks for using the antibiotics against wild-type strains. Those complications do not arise with the aforementioned (simple) refunding schemes in the environmental sciences. Third, one might also achieve sufficiently-strong incentives to develop new antibiotics without a refunding scheme by allowing for very high prices when an antibiotic is used against bacterial strains that are resistant against other antibiotics. We do not pursue this approach since enormously high prices for a treatment would raise ethical and health concerns, since certain therapies might then not be affordable anymore, which, in turn, would further fuel the spreading of resistant bacteria. Moreover, since the use of an antibiotic against wild-type strains exerts a negative externality on all individuals---due to the possible emergence of resistant bacteria in response to this use---all cases of antibiotics use contribute to the financing of narrow-spectrum antibiotics. Levying a fee on antibiotic use not only fills up the antibiotic fund and incentives the development of narrow-spectrum antibiotics, it also promotes the cautious use of existing antibiotics. Both effects internalize the negative externality caused by antibiotic use. 
\subsection{Refunding schemes for two antibiotics}
The mathematical model we formulate for the proposed refunding and incentivization scheme discussed above includes two elements:
\begin{itemize}

\item There is a fixed amount, denoted by $\alpha$, which a pharmaceutical company obtains if it successfully develops a new antibiotic $\mathrm{B}_i$, i.e.~an antibiotic approved by a public health authority.

\item There is a variable refund that is determined by the following refunding function:
\begin{equation}
    g(f_{1 \mathrm{B}_i}y_1, f_{3\mathrm{B}_i}y_3) = \beta \frac{f_{3\mathrm{B}_i}y_3}{\gamma f_{1\mathrm{B}_i}y_1+f_{3\mathrm{B}_i}y_3}\,,
\end{equation}
where $i\in \{1,2\}$ (to represents antibiotics $\mathrm{B}_1$ and $\mathrm{B}_2$) and $\beta$ and $\gamma$ are scaling parameters, with $\beta$ being a large number and $\gamma$ satisfying $\gamma \geq 1$. The refunding function $g(f_{1\mathrm{B}_i}y_1, f_{3\mathrm{B}_i}y_3)$ determines the relative use of the new antibiotic in compartment 3 (A-resistant strains) compared to the total use of the antibiotic weighted by the parameter $\gamma$. We require the refunding function to be
\begin{itemize}
    \item bounded according to $0 \leq g(f_{1\mathrm{B}_i}y_1, f_{3\mathrm{B}_i}y_3)  \leq \beta$,
    
    \item increasing in the use for currently resistant bacteria in comparison with other newly developed antibiotics used for this purpose: $f_{3\mathrm{B}_i}y_3$,
    \item declining in the use of antibiotics for non-resistant bacteria in comparison with other antibiotics used for this purpose: $f_{1\mathrm{B}_i}y_1$,
    \item maximal if the antibiotic is only used to treat A-resistant strains and 0 if it is only used for  nonresistant strain treatment.
\end{itemize}

\end{itemize}
Note that refunding scheme uses three free parameters $\alpha$, $\beta$, and $\gamma$. While in the simplest cases, only one or two parameters would be needed, we will see in the subsequent section that all three parameters are necessary to achieve the objective of the refunding scheme.  

The total refund that a successful pharmaceutical company receives in the time interval $[0,T]$ for developing an antibiotic $\mathrm{B}_i$ is given by 

\begin{equation}
R_i(T) \coloneqq \alpha +\int_0^{T} \beta \frac{f_{3\mathrm{B}_i}y_3 (f_{1\mathrm{B}_i}y_1+f_{3\mathrm{B}_i}y_3)}{\gamma f_{1\mathrm{B}_i}y_1+f_{3\mathrm{B}_i}y_3}\, \mathrm{d}t\,.
\label{eq:ref}
\end{equation}

We note that for $\gamma=1$, the refund is solely determined by $f_{3 \mathrm{B}}y_3$ and the use for compartment 1 is irrelevant for the refund. For $\gamma>1$, the use of antibiotics for compartment 1 decreases the refund.
\subsection{Incentivizing development}
We next focus on the research and development efforts of pharmaceutical companies and on how antibiotics are used once they have been developed. For this purpose, we first consider the situation without refunding. Companies are assumed to make a risk-neutral evaluation of the profit opportunities and loss risks from investing in such developments. For simplicity, we neglect discounting. Then, without refunding (i.e., without $R_i(T)$), the net profit of a company that develops an antibiotic $B_i$ is given by: 

\begin{equation}
    \pi_i = q_i (p_i-v_i) \int_0^T \frac{\mathrm{d} C_{\mathrm{B}_i}}{\mathrm{d}t}\, \mathrm{d}t - K_i,
        = q_i (p_i-v_i) \int_0^{T} (f_{1 \mathrm{B}_i}y_1+f_{3\mathrm{B}_i}y_3)\, \mathrm{d}t - K_i\,,
\end{equation}
where $K_i$ denotes the total development costs of $\mathrm{B}_i$, and $q_i$ the probability of success when the development is undertaken. Moreover, $p_i$ is the revenue per unit of the antibiotic used in medical treatments and $v_i$ are the production costs per unit. 

Note that in our example with two antibiotics, $f_{3 \mathrm{B}_i}=1$, since only drug $\mathrm{B}_i$ can be used for A-resistant strains. We assume that without refunding, $\pi_i$ is (strongly) negative, because of high development costs $K_i$ and low success probabilities $q_i$. The task of a refunding scheme is now three-fold: First, it has to render developing new antibiotics commercially viable. Second, it has to render developing narrow-spectrum antibiotics more attractive than developing broad-spectrum antibiotics. Third, if a narrow-spectrum antibiotic is developed that is also effective against wild-type strains, but less so than others, the refunding scheme should make the use against wild-type strains unattractive.

With a refunding scheme in place, we directly look at the conditions for such a scheme to achieve the break-even condition, i.e.~a situation at which $\pi$ becomes zero and investing into antibiotics development becomes just commercially viable. The general break-even condition for a newly developed antibiotic $\mathrm{B}_i$ is given by 
\begin{equation}
\label{eq:BE-funding}
\begin{aligned}
    K_i &= q_i (p_i-v_i) \int_0^{T} (f_{1\mathrm{B}_i}y_1+f_{3\mathrm{B}_i}y_3) \, \mathrm{d}t + q_i R_i(T)\\
        &= \alpha q_i + q_i\int_0^{T} \left[\beta\frac{f_{3\mathrm{B}_i}y_3}{\gamma f_{1\mathrm{B}_i}y_1+f_{3\mathrm{B}_i}y_3} + (p_i-v_i)\right] (f_{1\mathrm{B}_i}y_1+f_{3\mathrm{B}_i}y_3) \, \mathrm{d}t\,.
\end{aligned}
\end{equation}
Clearly, refunding increases the profits from developing new antibiotics. There are many combinations of the refunding parameters $\alpha$, $\beta$, and $\gamma$ that can achieve this break-even condition.  
However and more subtly, the refunding has to increase the incentives for the development of narrow-spectrum antibiotics more than those for broad-spectrum antibiotics. This can be achieved by an appropriate choice of the scaling parameter, as we will illustrate next. 
\subsection{Critical conditions for refunding parameters}
To derive the critical refunding parameters, we assume that the parameter $\alpha$, with $0< \alpha < K_i$, is given and thus a fixed share of the R\&D costs is covered by the antibiotics fund.
Based on the break-even condition (see Eq.~\eqref{eq:BE-funding}), we obtain the following general condition that the parameters $\beta$ and $\gamma$ have to satisfy:
\begin{equation}
\label{eq:b_crit}
    \beta = \frac{K_i-\alpha q_i-q_i(p_i-v_i)\int_0^T(f_{1 \mathrm{B}}y_1 + f_{3 \mathrm{B}}y_3)\mathrm{d}t}{q_i\int_0^T \frac{f_{3 \mathrm{B}}y_3}{\gamma f_{1 \mathrm{B}}y_1 + f_{3 \mathrm{B}}y_3}(f_{1 \mathrm{B}}y_1+f_{3 \mathrm{B}}y_3)\mathrm{d}t}.
\end{equation}
The goal of our refunding scheme is to incentivize pharmaceutical companies to produce narrow-spectrum antibiotics $\mathrm{B}_2$ that are only used against currently resistant strains (see treatment strategy IV in Sec.~\ref{sec:narrow_broad}). Thus, the refunding scheme has to satisfy two conditions: first, with the development of the antibiotic $\mathrm{B}_2$, the company achieves break-even. Second, developing antibiotic $\mathrm{B}_1$ is not attractive, i.e.~the profit is negative. To satisfy the first condition, we use Eq.~\eqref{eq:b_crit} and obtain the optimal value
\begin{equation}
\label{eq:b_ast_crit}
    \beta^\ast = \frac{K_2-\alpha q_2-q_2(p_2-v_2)\int_0^Tf_{3 \mathrm{B}_2}y_3 \, \mathrm{d}t}{q_2\int_0^T f_{3 \mathrm{B}_2}y_3 \, \mathrm{d}t}\,,
\end{equation}
where we used that $f_{1 \mathrm{B}_2}=0$ (see Sec.~\ref{sec:narrow_broad}).
To achieve negative profit for using $\mathrm{B}_1$, we need to choose the parameter $\gamma$ such that developing a broad-spectrum antibiotic $\mathrm{B}_1$ and applying it in $y_1$ and $y_3$ (see treatment strategy I in Sec.~\ref{sec:narrow_broad}) is not more attractive than developing a narrow-spectrum antibiotic $\mathrm{B}_2$ according to treatment strategy IV (see Sec.~\ref{sec:narrow_broad}). Thus, the refunding scheme needs to satisfy
\begin{equation}
\label{eq:Condition_for_c}
\alpha q_1 + q_1\int_0^{T} \left[\beta^\ast\frac{f_{3\mathrm{B}_1}y_3}{\gamma f_{1\mathrm{B}_1}y_1+f_{3\mathrm{B}_1}y_3} + (p_1-v_1)\right] (f_{1\mathrm{B}_1}y_1+f_{3\mathrm{B}_1}y_3) \, \mathrm{d}t - K_1 < 0\,.
\end{equation}
If we evaluate inequality \ref{eq:Condition_for_c} as an equality, we obtain a critical value for $\gamma$, denoted by $\gamma_1^*$, for certain values of $\beta^*$, $p_1$, and $v_1$. For $\gamma>\gamma_1^*$, it is more profitable to produce a narrow-spectrum antibiotic $\mathrm{B}_2$ and get a higher refund than to develop a broad-spectrum antibiotic $\mathrm{B}_1$ and sell more units. We observe that this critical value is uniquely determined, since the left side is strictly decreasing in $\gamma$. We discuss conditions for the existence of $\gamma_1^*$ in the next section.

Third, we need to make sure that a narrow-spectrum antibiotic $\mathrm{B}_2$ is not used for wild-type bacterial strains (see the 50/50 treatment strategy III in Sec.~\ref{sec:narrow_broad}). Since a narrow-spectrum antibiotic may be also effective for wild-type strains, the refunding scheme should exclude any incentives to use $\mathrm{B}_2$ in compartment $Y_1$. In terms of our refunding scheme, this could be achieved by replacing the 50/50 treatment strategy involving antibiotic $\mathrm{B}_1$ on the left-hand side of Eq.~\eqref{eq:Condition_for_c} with the 50/50 treatment strategy involving antibiotic $\mathrm{B}_2$. Note that the resulting critical value for $\gamma$, which we denote by $\gamma_2^\ast$, is different from $\gamma_1^*$. An alternative to imposing this additional constraint on the refunding scheme is to implement strict medical guidelines which demand that less-effective antibiotics should not be used in compartment $Y_1$.

Together, Eqs.~\eqref{eq:b_ast_crit} and \eqref{eq:Condition_for_c} determine the refunding scheme that ensures that a pharmaceutical company breaks even at time $T$ by developing a narrow-spectrum antibiotic, does not focus on broad-spectrum antibiotics and the narrow-spectrum antibiotics is not misused once it is developed.  
\subsection{Numerical example}
\begin{figure}
    \centering
    \includegraphics[width = \textwidth]{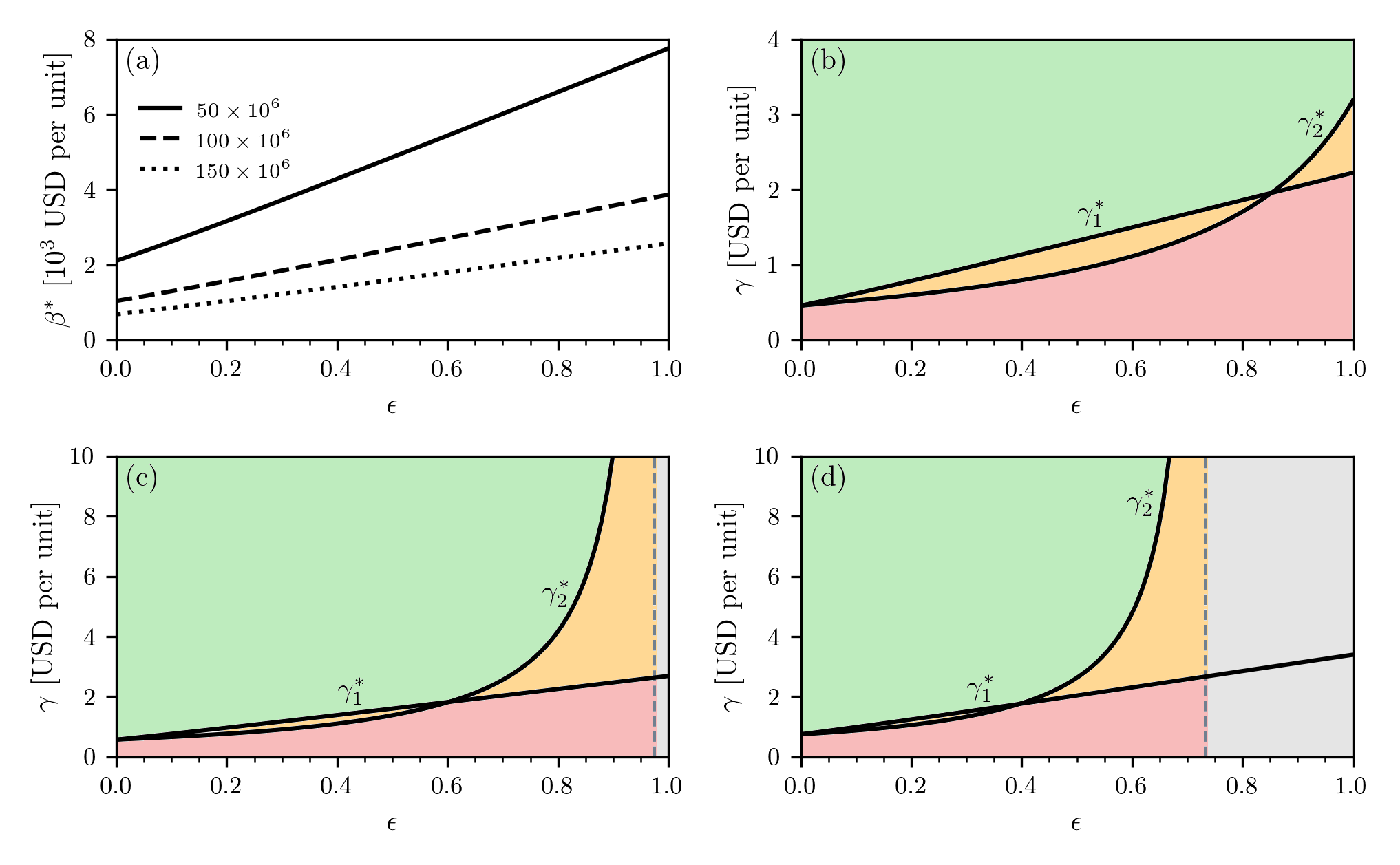}
    \caption{\textbf{Critical refunding parameters.} We show the critical refunding parameters $\beta^*$ (a) and $\gamma^*$ (b--d) for different $\epsilon$ and population sizes $50\times 10^6$, $100\times 10^6$, and $150\times 10^6$. The critical refunding parameter $\beta^*$ increases with population size (a). A finite $\gamma_1^*$ and $\gamma_2^*$ is indicated by the black solid line, whereas the left hand side of Eq.~\eqref{eq:Condition_for_c} is always positive in the grey-shaded region (b--d). We numerically solve Eqs.~\eqref{eq:narrow_broad_new} with a classical Runge-Kutta scheme in the time interval $[0,T]$ with $T=100$ setting $\lambda=100$, $d=1$, $c=1.5$, $b=0.03$, $r_{i}=(2-k) 0.1$ ($k$ is the number of effective antibiotics in the respective layer), $h=1$, $s=0.05$, $\alpha_\mathrm{B}=\gamma_{\mathrm{B}}=\epsilon \geq 0$. The initial conditions are $x(0)=50$, $y_1(0)=33.33$, $y_2(0)=y_3(0)=y_4(0)=0$. All compartments were rescaled according to the shown population sizes.}
    \label{fig:numerical_parameters}
\end{figure}
\begin{table}
\begin{tabular}{l|l|l}
 \textbf{quantity} &  \textbf{symbol}   & \textbf{value}      \\ \hline
success probability & $q_i$   & 0.1        \\
development costs & $K_i$   & 2 billion USD\\
refunding offset & $\alpha$   & 1 billion USD\\
revenue per unit & $p_i$   & 100 USD  \\
production costs per unit & $v_i$   &  70 USD \\
\end{tabular}
\caption{Refunding calibration parameters.}
\label{tab:parameters}
\end{table}

We now focus on a simple example to illustrate how our refunding scheme can incentivize the development of narrow-spectrum antibiotics. For this purpose, we use the parameters listed in Tab.~\ref{tab:parameters}. To work with reasonable population sizes, we apply our refunding scheme to populations with 50, 100, and 150 million people and rescale the corresponding compartments that we used to determine the antibiotic consumption in Fig.~\ref{fig:consumption}.

We first determine $\beta^*$ according to Eq.~\eqref{eq:b_ast_crit} and show the results in Fig.~\ref{fig:numerical_parameters} (a). Since the consumption $C_{\mathrm{B}_2}$ decreases with $\epsilon$ (see Fig.~\ref{fig:consumption}), the critical refunding parameter $\beta^*$ has to increase with $\epsilon$. Before discussing the corresponding values of $\gamma_1^*$ and $\gamma_2^*$, we briefly summarize the conditions for the existence of a critical value $\gamma^*$ and distinguish three cases.
\begin{itemize}
\item[Case I:] If $q_i a+ q_i \int_{0}^T \left[\beta^*+(p_i-v_i) \right] (f_{1\mathrm{B}_i}y_1+f_{3\mathrm{B}_i}y_3) \mathrm{d}t - K_i< 0$ ($i=1,2$ and $f_{(\cdot)}$ is chosen according to some treatment strategy), we find that Eq.~\eqref{eq:Condition_for_c} is satisfied for any $\gamma > 0$ independent of the underlying refunding scheme since, for finite $\gamma$, 
\begin{equation}
\beta^\ast\frac{f_{3\mathrm{B}_i}y_3}{\gamma f_{1\mathrm{B}_i}y_1+f_{3\mathrm{B}_i}y_3}=\beta^\ast \frac{1}{1 + \gamma \frac{f_{1\mathrm{B}_i}y_1}{f_{3\mathrm{B}_i}y_3}} \leq \beta^\ast\,.
\end{equation}

\item[Case II:] If $q_i a+ q_i \int_{0}^T \left[\beta^*+(p_i-v_i) \right] (f_{1\mathrm{B}_i}y_1+f_{3\mathrm{B}_i}y_3) \mathrm{d}t - K_i > 0$ and $q_i a+ q_i \int_{0}^T (p_i-v_i) (f_{1\mathrm{B}_i}y_1+f_{3\mathrm{B}_i}y_3) \mathrm{d}t - K_i < 0$, there exists a $\gamma^* > 0$ such that the left-hand side of Eq.~\eqref{eq:Condition_for_c} (for $\mathrm{B}_i$ and corresponding refunding parameters) is equal to zero.

\item[Case III:]  If $q_i a+ q_i \int_{0}^T (p_i-v_i) (f_{1\mathrm{B}_i}y_1+f_{3\mathrm{B}_i}y_3) \mathrm{d}t - K_i > 0$, it is not possible to satisfy Eq.~\eqref{eq:Condition_for_c} (for $\mathrm{B}_i$ and corresponding refunding parameters), since $p_i-v_i$ is too large. 
\end{itemize}
For the parameters of Tab.~\ref{tab:parameters}, we show the resulting $\gamma_1^*$ and $\gamma_2^*$ in Fig.~\ref{fig:numerical_parameters} (b--d). We observe that $\gamma_1^*$ always exists for the chosen parameters, whereas $\gamma_2^*$ only exists for certain values of $\epsilon$ (case II). Case I does not exist in the outlined example, since the chosen $\beta^*$ (Eq.~\eqref{eq:b_ast_crit}) is too large to satisfy Eq.~\eqref{eq:Condition_for_c} (evaluated for $\mathrm{B}_2$ and corresponding refunding parameters) for any $\gamma > 0$. At the boundary separating cases II and III, we find that $\gamma_2^*$ diverges. Note that the size of region associated with case III increases with population size. To avoid such scenarios in real-world applications of our refunding scheme, we could, for instance, reduce the refunding offset $\alpha$.

To summarize:
\begin{itemize}
    \item For intermediate consumption of $\mathrm{B}_2$ in 50/50 treatment (see treatment strategy III in Sec.~\ref{sec:narrow_broad}) and corresponding returns, a finite $\gamma_2^*$ exists (see Fig.~\ref{fig:numerical_parameters} (b--d)). Within the \textcolor{green1}{green-shaded regions} of Fig.~\ref{fig:numerical_parameters} (b--d), Eq.~\eqref{eq:Condition_for_c} is satisfied for $\mathrm{B}_1$ and $\mathrm{B}_2$ ($\gamma > \gamma_1^*$ and $\gamma > \gamma_2^*$), whereas the left-hand side of Eq.~\eqref{eq:Condition_for_c} is positive for $\mathrm{B}_1$ and $\mathrm{B}_2$ within the \textcolor{red1}{red-shaded regions} ($\gamma < \gamma_1^*$ and $\gamma < \gamma_2^*$).
    \item Within the \textcolor{orange1}{orange-shaded regions} of Fig.~\ref{fig:numerical_parameters} (b--d), either $\gamma > \gamma_1^*$ or $\gamma > \gamma_2^*$.
    \item If the expected return associated with the $\mathrm{B}_2$ treatment strategy III of Sec.~\ref{sec:narrow_broad} is too large, there is no $\gamma>0$ that discourages pharmaceuticals from developing such drugs (\textcolor{grey1}{grey-shaded regions} in Fig.~\ref{fig:numerical_parameters} (c) and (d)). 
\end{itemize}
\section{Refunding schemes: General considerations}
\label{sec:refunding_general}
Using a refunding scheme as constructed above in practice requires a series of additional considerations which we address in this section. In particular, it must be possible to apply a refunding scheme for any constellation of antibiotics use, it must work under a variety of sources of R\&D uncertainty, and it must still be effective when diagnostic and treatment uncertainties are taken into account.

\subsection{Generalizations}
We first generalize the refunding scheme for possible treatment with more than 2 antibiotics. We assume that currently, $N_1$ antibiotics are used. Now, $N_2$ new antibiotics are potentially developed, such that the total number of antibiotics is $N=N_1+N_2$. Note that before new antibiotics are introduced, there is 1 non-resistant strain and a total of $2^{N_1}-1$ strains that are resistant to some antibiotic. Furthermore, there is one class of bacterial strains that is resistant to all antibiotics currently on the market. The class of microbes that is resistant to all $N$ antibiotics has the index $\hat{k}=2^N$. The generalized refunding scheme still consists of a fixed refund $\alpha$ and a variable refund that depends on the use of the antibiotic in the different compartments. The scaling parameter $\gamma_1$ "punishes" the use of the antibiotic for wild-type strains by decreasing the refund. In addition, $\gamma_2$ scales the reward of the use of the antibiotic for strains that are resistant to some, but not all, antibiotics currently on the market. Note that $\gamma_2$ could be negative, such that the refund still increases in the use for partially resistant strains. Lastly, the refund strongly increases in the use for fully-resistant strains in the class $\hat{k}=2^N$. 

The refunding scheme is given by:

\begin{equation}
   g(\mathbf{\vec{f}_i}) = \beta\frac{\sum_{j=2}^{2^N}f_{j \mathrm{B}_i}y_j}{\gamma_1f_{1 \mathrm{B}_i}y_1+\gamma_2\sum_{j=2}^{2^N-1}f_{j \mathrm{B}_i}y_j + f_{\hat{k} \mathrm{B}_i}y_{\hat{k}}}\,,
\end{equation}

where $\mathbf{\vec{f}_i}$ denotes the vector of the usage of a newly developed drug $\mathrm{B}_i$ in all compartments $Y_j$ with $j \in \{1,\dots,N\}$. The use in each compartment is given by $f_{j \mathrm{B}_i}y_j$.

The break-even conditions can be established as for the 2-antibiotics case, but now with adjusted total consumption per antibiotic and with the generalized refunding scheme. 

Similarly to the extension to more than 2 antibiotics, the refunding scheme can be generalized when more than one pharmaceutical company should be given incentives to do R\&D on narrow-spectrum antibiotics. In such cases, the refunding parameters have to be adjusted, such that with lower sales volumes for each company, it is still profitable to undertake the R\&D investments.
\subsection{Multi-dimensional R\&D uncertainties}
The development and usage of antibiotics is subject to a variety of uncertainties. In particular, companies may not know at the beginning of a development process against which type of bacterial strains the drug that might emerge will be effective. Such uncertainties can be taken into account as follows:    
Suppose a pharmaceutical company starts an R\&D process for an antibiotic, but does not know initially, whether it will turn out to be broad or narrow-spectrum, as this will only become known during or, in the worst case, at the end of the development process. 

A possible solution to this issue is setting the value of $\beta$ equal to  $\beta^\ast + \delta$ for some small $\delta >0$. Moreover, we set $\gamma$ at the critical value $\gamma^*$ for the value of $\beta^\ast+ \delta$. Then, starting the R\&D investment is profitable and the incentives for a narrow-spectrum antibiotic are maximal. If during the R\&D process, a narrow-spectrum opportunity emerges, it will be chosen, since profits will be higher than for a broad-spectrum antibiotic. However, the company also breaks even for a broad-spectrum antibiotic. Hence, the company faces no additional risk if it is impossible at the beginning to evaluate whether a broad or narrow-spectrum antibiotics will result from the R\&D investment. 

\subsection{Diagnostics and treatment uncertainties}

The refunding scheme depends on the ability of doctors to rapidly identify the strain of bacteria that caused a certain infection. For a fraction of such treatments, this may be impossible - in particular in emergency situations or when rapid, high-throughput diagnostic devices are unavailable. Note that certain bacterial strains can already be identified in a few hours by using peptide nucleic acid (PNA) fluorescent in-situ hybridization (FISH) tests, mass spectroscopy, and polymerase chain reaction (PCR)-based methods~\cite{kothari2014emerging}.

The refunding scheme can be readily adapted to allow for diagnostic and treatment uncertainties. For instance, one could base refunding only on diagnosed strains of bacteria against which the antibiotics is used. The refunding parameters have to be adapted accordingly.
Basing refunding only on cases in which the bacterial strain has been diagnosed and reported would provide further incentives for pharmaceuticals to develop fast diagnostic tests to identify the sources of infections.

A further refinement would be to provide a refund in case a newly developed antibiotic is used and turns out to be effective. Such success targeting would be desirable, but may not be easily implementable in practice. As long as the success rates of an antibiotic that is effective against particular bacterial strains are known or can be estimated with sufficient precision, taking the usage/bacterial strain data would be sufficient to provide desirable incentives to engage in R\&D for narrow-spectrum antibiotics.
\subsection{Small firms and the R\&D ecosystem}
Both small biotech companies and large pharmaceutical companies play a significant role in developing new antibiotics. The flexibility, nimbleness, and flat organizational structure of smaller biotech companies that specialize in innovative antibacterial treatments can be very effective for the development of new antibiotics. Therefore, while refunding will mostly benefit large pharmaceutical companies, the anticipation of such refunds is expected to also motivate smaller biotech companies to step up with R\&D efforts. These smaller companies can expect significant rewards when they sell or license their patents to larger companies. Moreover, small biotech companies may receive much more start-up funding both from venture capitalists and larger pharmaceutical companies, and one might even consider using the antibiotics fund for this purpose as well. Hence, it is expected that the refunding scheme will be nourishing for the entire ecosystem that develops new antibiotics. 
\subsection{The antibiotics fund and participating countries}
A necessary condition for the functioning of our refunding scheme is the existence of an antibiotics fund with sufficient equity to cover R\&D incentives. Similar to the recently established AMR Action Fund, which aims at bridging the gap between the pipeline for innovative antibiotics and patients, an antibiotics fund should be started by initial contributions from industry and public institutions. Since it is in the collective self-interest of the pharmaceutical industry to solve the antibiotics dilemma---as otherwise many other business lines and their reputation will be harmed---a significant contribution from the industry to set up the antibiotics fund can be expected, as was the case for the AMR Action Fund.  

In addition, a continuous refilling of the fund can be achieved by levying a fee on every use of existing antibiotics. These fees have to be set in such a way that the antibiotics fund will never be empty. Since the (sometimes excessive) use of existing antibiotics (e.g., in agricultural settings~\cite{swann_report,casey2013high,xu2020antibiotic}) creates the resistance problem, levying this fee not only helps to continuously refill the fund, but it may also help to use existing antibiotics cautiously. Ultimately, the antibiotics fund and the refunding scheme are a mechanism to internalize the externality in antibiotics use, namely the creation of resistant bacteria.

As in the context of slowing down climate change, the ideal implementation would involve a global refunding scheme administered by an international agency, because reducing resistance is a global public good. However, also similar to implementing climate-change policies, worldwide adoption should be extremely difficult and might be impossible to achieve. As a starting point, a set of industrialized countries should agree to a treaty that fails if any of them
does not participate. Once an antibiotic fund has been initiated, a treaty should establish the continuing financing of the antibiotics fund and the refunding scheme. The gains would be large and may lead to long-standing self-enforcing
incentives to substantially and continuously
increase the chances to develop antibiotics against resistant bacteria. If attempts to build a larger coalition fail, the European Union or the US could take the lead and become the first country or coalition of countries that implements a refunding scheme for antibiotics.

\section{Conclusions}
\label{sec:conclusion}
The rapid rise of antibiotic resistance poses a serious threat to global public health. No new antibiotic has been registered or patented for more than three decades (see Fig.~\ref{fig:antibiotic}). To counteract this situation, we introduced a novel framework to mathematically describe the emergence of antibiotic resistance in a population that is treated with $n$ antibiotics. We then used this framework to develop a market-based refunding scheme that can solve the antibiotics dilemma, i.e.~which can incentivize pharmaceutical companies to reallocate resources to antimicrobial drug discovery and, in particular, to the development of narrow-spectrum antibiotics that are effective against multiresistant bacterial strains. We outlined how such a refunding scheme can cope with various sources of uncertainty inherent to R\&D for antibiotics as well as with diagnostic and treatment uncertainties.
\acknowledgments{LB acknowledges financial support from the SNF Early Postdoc.Mobility fellowship on ``Multispecies interacting stochastic systems in biology'' and the US Army Research Office (W911NF-18-1-0345). We thank Margrit Buser, Emma Schepers, and Maria R.~D'Orsogna for helpful comments and corrections. LB also acknowledges helpful discussions with Paul Richter.
}

\newpage
\appendix
\renewcommand\thefigure{\thesection.\arabic{figure}}
\setcounter{figure}{0}    
\section{Combination therapy versus targeted use of antibiotics}
\label{app:comparison}
\begin{figure}
\centering
\includegraphics[width=0.49\textwidth]{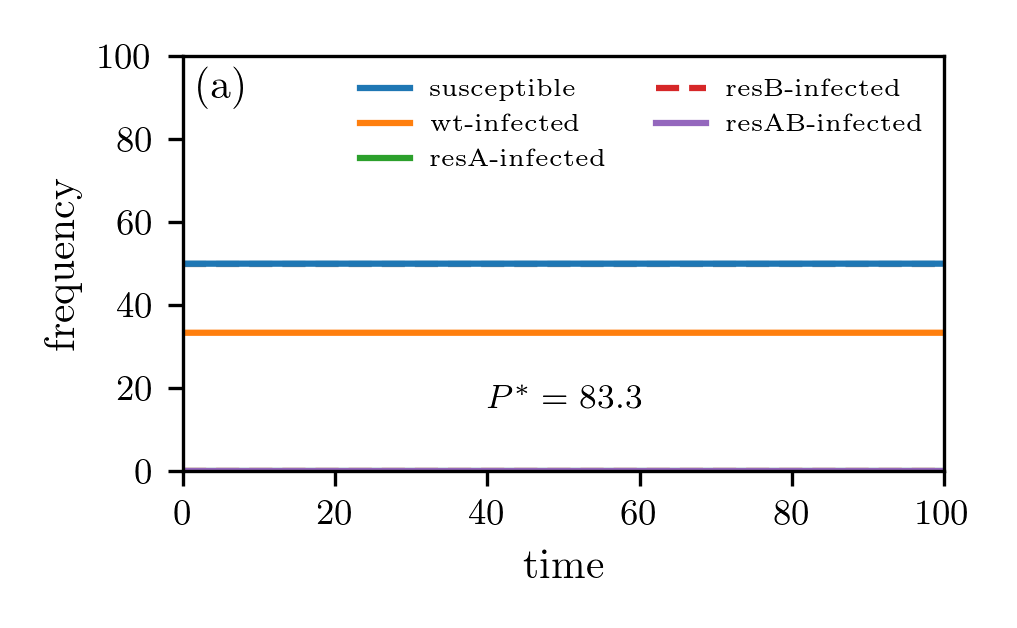}
\includegraphics[width=0.49\textwidth]{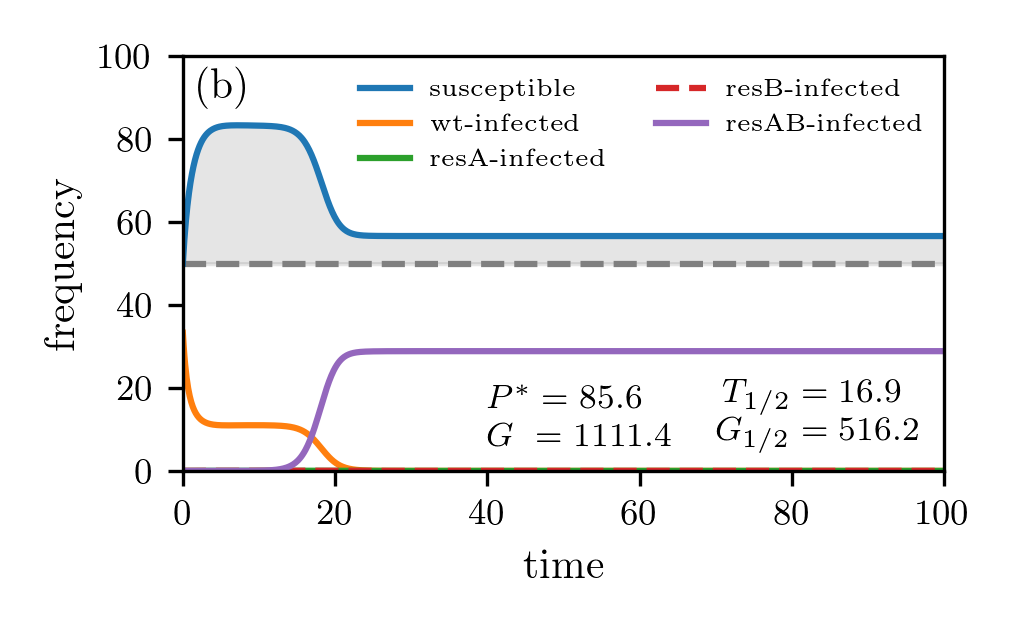}
\includegraphics[width=0.49\textwidth]{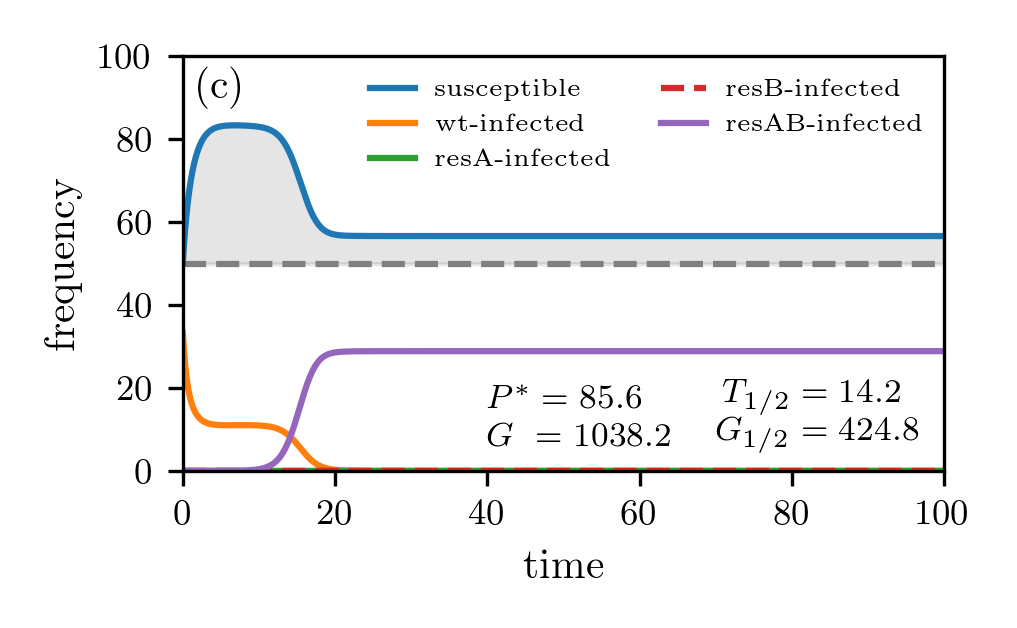}
\includegraphics[width=0.49\textwidth]{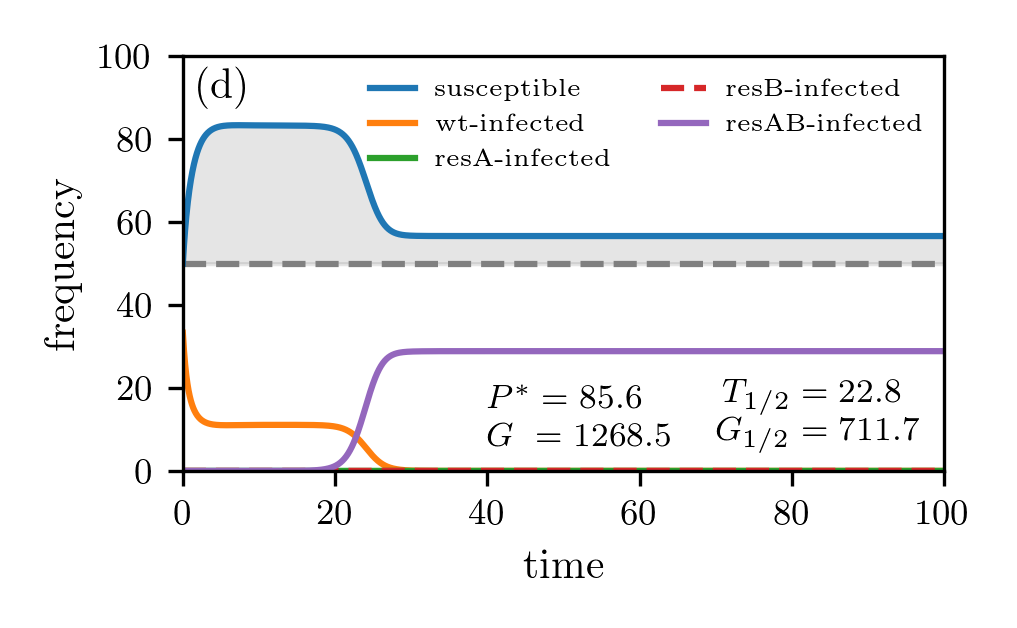}
\caption{\textbf{Treatment with two antibiotics.} We numerically solve Eqs.~\eqref{eq:n2_antibiotics1} and \eqref{eq:n2_antibiotics2} with a classical Runge-Kutta scheme setting $\lambda=100$, $d=1$, $c=1.5$, $b=0.03$, $r_{1}=0$, $r_{2}=r_{3}=0.1$, $r_{4}=0.2$. In panel (a), we set $h=0$ (i.e., no treatment) and $h=1$ in the remaining panels. Panel (b) shows a solution of Eq.~\eqref{eq:n2_antibiotics1} (``single-antibiotic therapy''). The bottom panels show solutions of Eq.~\eqref{eq:n2_antibiotics2} (``combination therapy'') with $q=10^{-5}>s^2=10^{-6}$ (c) and $q=10^{-8}<s^2=10^{-6}$ (d). If $q>s^2$, the gain is smaller for multiple treatment. We use $P^\ast$ and $G$ to indicate the total stationary population size (see Eqs.~\eqref{eq:stat_pop} and \eqref{eq:pop_param}) and gain of uninfected in the considered time interval (see Eq.~\eqref{eq:gain}), respectively. The gain $G$ corresponds to the grey shaded region and the characteristic resistance time scale $T_{1/2}$ is the time when the proportion of completely resistant strains is 50\% (see Eq.~\eqref{eq:T12}). $G_{1/2}$ is the gain in the time interval $[0,T_{1/2}]$. The initial conditions are $x(0)=50$, $y_1(0)=33.33$, and $y_2(0)=y_3(0)=y_4(0)=0$.}
\label{fig:simulations1}
\end{figure}
To illustrate the difference between combination- and single-antibiotic therapy, we first derive the corresponding mathematical results for $n=2$ antibiotics $\{\mathrm{A},\mathrm{B}\}$~\cite{bonhoeffer1997evaluating} and discuss the case where $n>2$ in Appendix \ref{app:general}.

For $n=2$ antibiotics, the corresponding sets of antibiotics for the $N=4$ infected compartments are $A_1=\{\mathrm{A},\mathrm{B}\}$, $A_2=\{\mathrm{A}\}$, $A_3=\{\mathrm{B}\}$, and $A_4=\emptyset$. Based on Eq.~\eqref{eq:rate1}, the treatment of patients with single broad-spectrum antibiotics can be described by:
\begin{align}
\begin{split}
\frac{\mathrm{d} x}{\mathrm{d} t}&=-b x \left( y_1+y_2+y_3+y_4\right)+r_1 y_1+r_2 y_2 + r_3 y_3+r_4 y_4 \\
&+  h (1-s) \left(y_1+y_2+y_3\right)+\lambda - d x\,,\\
\frac{\mathrm{d} y_1}{\mathrm{d} t}&=\left(b x - r_1 -  h-c \right)y_1\,, \\
\frac{\mathrm{d} y_2}{\mathrm{d} t}&=\left(b x  - r_2 - h-c \right)y_2+\frac{1}{2}h s y_1\,, \\
\frac{\mathrm{d} y_3}{\mathrm{d} t}&=\left(b x  - r_3- h-c \right)y_3+\frac{1}{2}h s y_1\,, \\
\frac{\mathrm{d} y_4}{\mathrm{d} t}&=\left(b x  - r_4-c \right)y_4 +h s \left( y_2 +y_3\right) \,,
\end{split}
\label{eq:n2_antibiotics1}
\end{align}
where we set $b_{ij}=b$, $s_{ij}=s$, $h_{ij}=h$, and use the proportions $f_{1 \mathrm{A}}=f_{1 \mathrm{B}}=1/2$, $f_{1 \mathrm{A B}}=0$, $f_{2 \mathrm{A}}=1$, $f_{2 \mathrm{B}}=f_{2 \mathrm{AB}}=0$, $f_{3 \mathrm{B}}=1$, and $f_{3 \mathrm{A}}=f_{3 \mathrm{AB}}=0$.

In the absence of treatment, the total stationary population is 
\begin{equation}
P^\ast=x^\ast+y_1^\ast=\frac{r_1+c}{b}+\frac{\lambda}{c}-\frac{d}{b}-\frac{d r_1}{c b}\,.
\label{eq:pop_param}
\end{equation}
Analytical expressions for $G$ and $T_{1/2}$ for some specific parameter constellations are summarized in Ref.~\cite{bonhoeffer1997evaluating}. We compare the single-antibiotic therapy and targeted use of antibiotics (see Eq.~\eqref{eq:n2_antibiotics1}) with a broad-spectrum treatment that uses combinations of antibiotics $\mathrm{A}$ and $\mathrm{B}$:

\begin{align}
\begin{split}
\frac{\mathrm{d} x}{\mathrm{d} t}&=-b x \left( y_1+y_2+y_3+y_4\right)+r_1 y_1+r_2 y_2 + r_3 y_3+r_4 y_4 \\
&+ h (1-q) y_1+ h (1-s) \left(y_2+y_3\right)+\lambda - d x\,,\\
\frac{\mathrm{d} y_1}{\mathrm{d} t}&=\left(b x  - r_1 -  h-c\right)y_1\,, \\
\frac{\mathrm{d} y_2}{\mathrm{d} t}&=\left(b x - r_2 - h -c\right)y_2\,, \\
\frac{\mathrm{d} y_3}{\mathrm{d} t}&=\left( b x - r_3 - h-c \right)y_3\,, \\
\frac{\mathrm{d} y_4}{\mathrm{d} t}&=\left(b x - r_4-c\right) y_4 +h q y_1+h s \left( y_2 +y_3\right) \,,
\end{split}
\label{eq:n2_antibiotics2}
\end{align}
where $q$ is the fraction of double resistances that develop from the combined treatment of the wild-type strain ($Y_1$) with antibiotics $\mathrm{A}$ and $\mathrm{B}$. In Eq.~\eqref{eq:n2_antibiotics2}, we use the proportions $f_{1 \mathrm{AB}}=1$, $f_{1 \mathrm{A}}=f_{1 \mathrm{B}}=0$, $f_{2 \mathrm{A B}}=1$, $f_{2 \mathrm{A}}=f_{2 \mathrm{B}}=0$, $f_{3 \mathrm{AB}}=1$, and $f_{3 \mathrm{A}}=f_{3 \mathrm{B}}=0$. 

We show a comparison between the outlined single-antibiotic and combination therapy treatment in Fig.~\ref{fig:simulations1}. If $q>s^2$, we find that, in agreement with earlier results~\cite{bonhoeffer1997evaluating}, the single-antibiotic treatment outperforms combination therapy. For $q < s^2$ (i.e., for very small probabilities of double resistance resulting from combination treatment of wild-type strains), single-antibiotic treatment is not as efficient as broad-spectrum therapy anymore.
\section{Properties of the general model}
\label{app:general}
\begin{figure}
\centering
\includegraphics[width=0.49\textwidth]{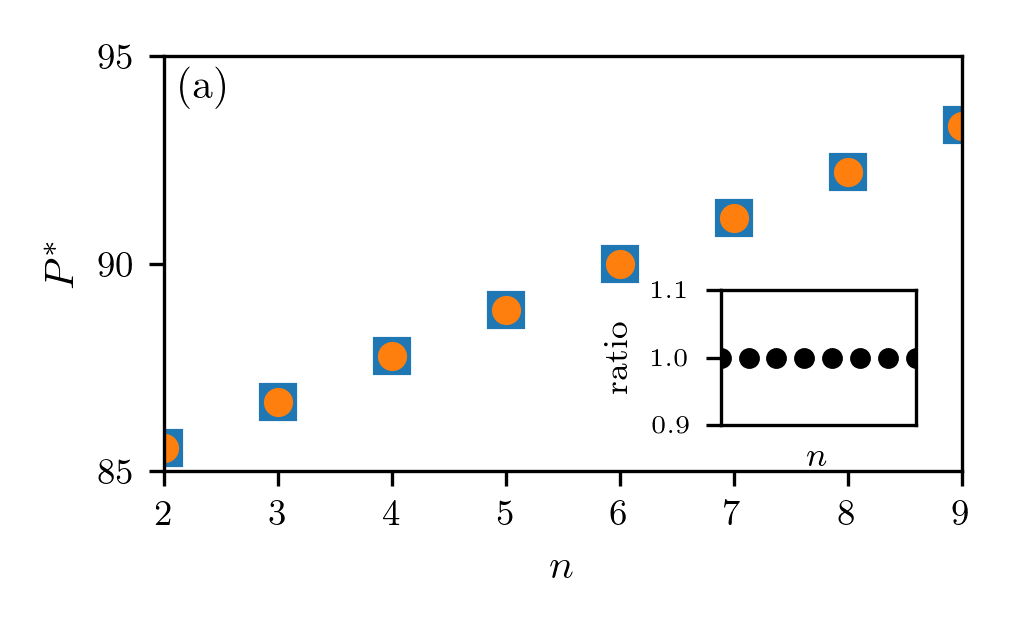}
\includegraphics[width=0.49\textwidth]{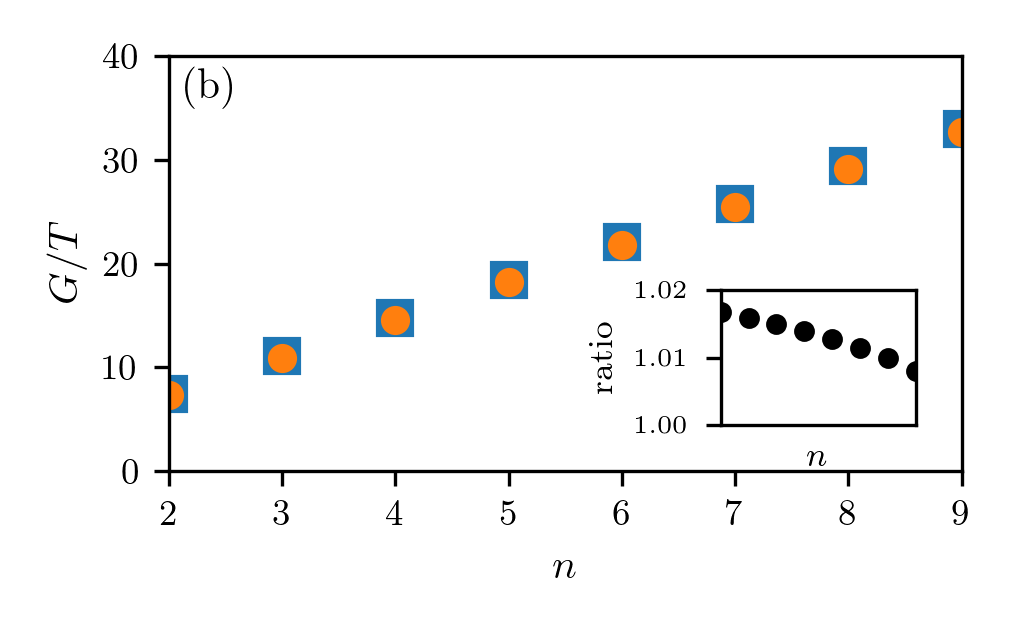}
\includegraphics[width=0.49\textwidth]{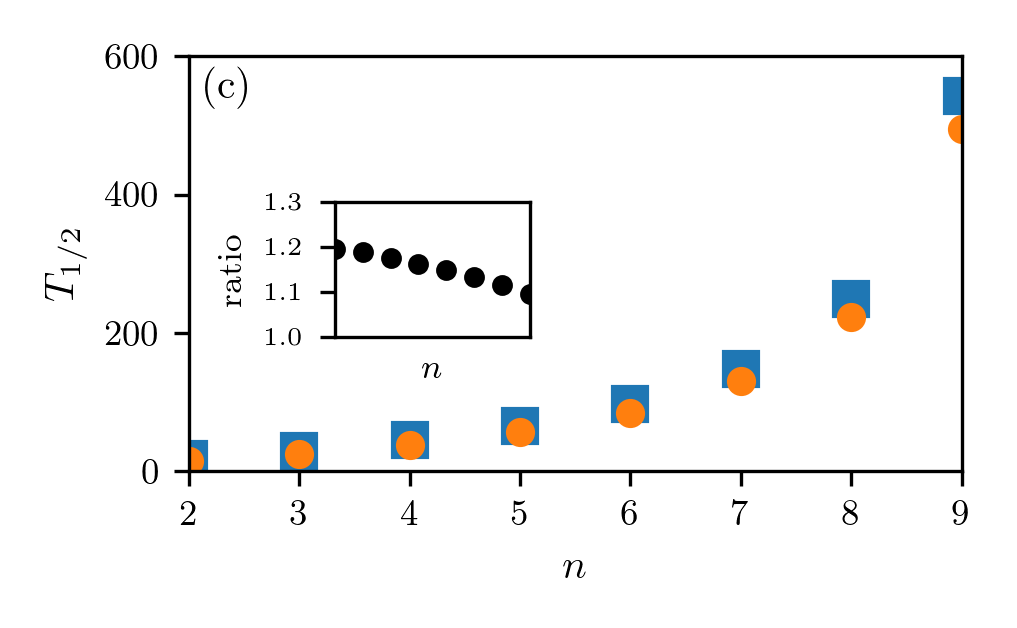}
\includegraphics[width=0.49\textwidth]{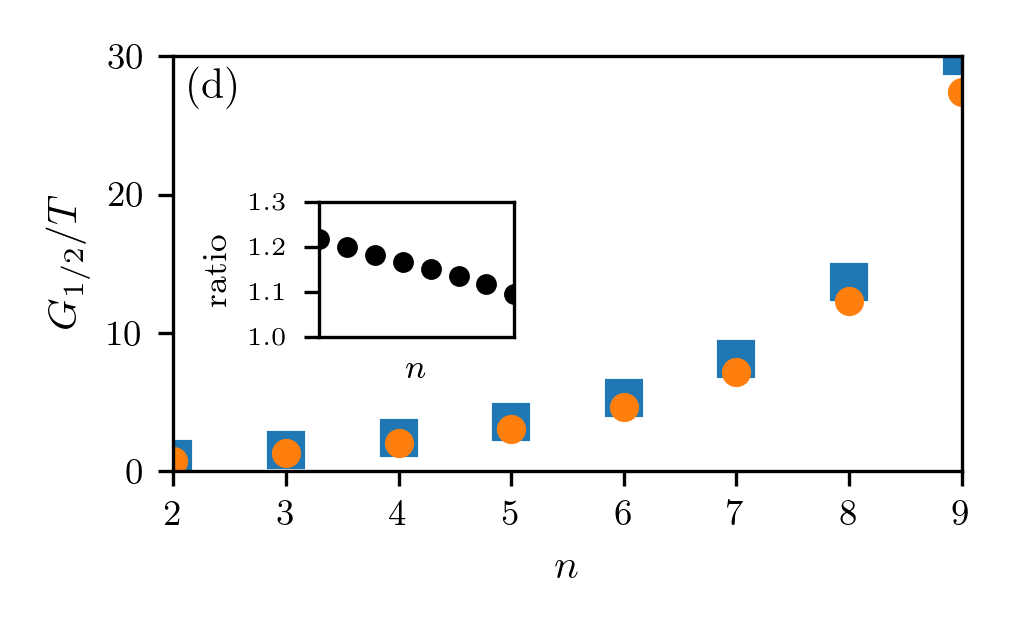}
\includegraphics[width=0.42\textwidth]{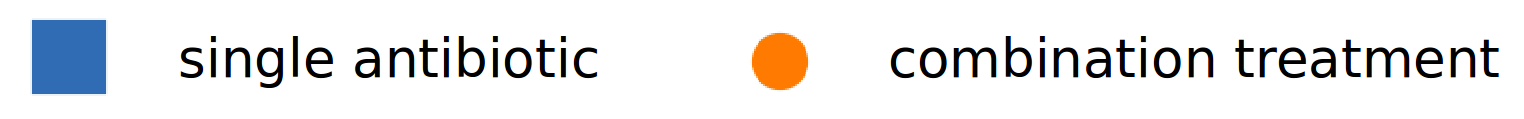}
\caption{\textbf{Treatment with multiple antibiotics.} We numerically solve the $n$-antibiotic generalizations of Eqs.~\eqref{eq:n2_antibiotics1} and \eqref{eq:n2_antibiotics2} with a classical Runge-Kutta scheme in the time interval $[0,T]$ with $T=600$ and $\lambda=100$, $d=1$, $c=1.5$, $b=0.03$, $r_{i}=(n-k) 0.1$ ($k$ is the number of effective antibiotics in the respective layer), $h=1$, $s=10^{-3}$, and $q=10^{-2.5 n}$. In panels (a) and (b), we show the total stationary population size $P^\ast$ (see Eqs.~\eqref{eq:stat_pop} and \eqref{eq:pop_param}) and the gain of uninfected $G$ (see Eq.~\eqref{eq:gain}); in panels (c) and (d), we show the time $T_{1/2}$ when the proportion of completely resistant strains is 50\% (see Eq.~\eqref{eq:T12}) and the gain $G_{1/2}$ in the time interval $[0,T_{1/2}]$. The insets in each panel represent the ratio of the single and combination therapy values. The initial conditions are $x(0)=50$, $y_1(0)=33.33$, and $y_2(0)=y_3(0)=y_4(0)=0$.}
\label{fig:simulations2}
\end{figure}
In this Appendix, we establish several properties of the general antibiotic-treatment model (see Eq.~\eqref{eq:rate1}). In particular, in the absence of antibiotic treatment and for sufficiently strong treatment, we show that the mathematical structure of the equation describing the stationary population $P^*$ is unaffected by the number of antibiotics and differences in treatment protocols. The term ``sufficiently strong treatment''~\cite{bonhoeffer1997evaluating} means that growth factor $b/(r_N+c_N)$ in the completely resistant compartment is larger than the growth factor $b/(r_i+c_i+\sum_{j\in A_i} f_{i j} h_{ij})$ in any other compartment $y_i$ ($i < N$).

In the absence of antibiotic therapy (i.e., $h_{ij}=0$ for all $i$, $j$), we find that the stationary population of susceptible individuals is $x^* = (r_1+c_1)/b_1$ and $y^* = \lambda/c_1 - d/b_1 - (d r_1)/(b_1 c_1)$. 

The stationary solution under sufficiently strong treatment with $y_N^\ast \neq 0$ implies that $y_1^\ast=y_2^\ast = \ldots = y_{N-1}^\ast =0$. Independent of treatment protocol details, the stationary solution is $x^\ast=(r_N+c_N)/b_N$ and $y_N^\ast = \lambda/c_N-d/b_N-(d r_N)/(b_N c_N)$. Even if the stationary behavior has a similar form for general numbers of antibiotics $n$, the dynamical features and corresponding characteristics such as $T_{1/2}$ exhibit a more complex dependence on $n$, which we analyze numerically in Fig.~\ref{fig:simulations2}.

We observe that for the considered parameters in the ``$q > s^n$ regime'' (see Appendix \ref{app:comparison}), single-antibiotic therapy treatment still outperforms combination treatment. The larger the number of antibiotics $n$, the smaller relative differences between various performance metrics (see insets in Fig.~\ref{fig:simulations2}).
\end{document}